\documentclass[amsmath,amssymb,superscriptaddress,aps,prl,reprint, 10pt]{revtex4-2}

\usepackage{layouts}

\usepackage[english]{babel}
\usepackage[utf8]{inputenc}
\usepackage[T1]{fontenc}                        

\usepackage{graphicx}
\usepackage{sidecap} 
\sidecaptionvpos{figure}{t}                        
\usepackage[section,below]{placeins}            
\usepackage{mwe}

\usepackage{tikz}
\usetikzlibrary{shapes.geometric, arrows}
\usetikzlibrary{decorations.markings}
\usetikzlibrary{decorations.pathmorphing}
\usetikzlibrary{decorations.pathreplacing}
\usetikzlibrary{shapes.symbols}

\usepackage{bm}
\usepackage{mathtools}                          
\usepackage{physics}
\allowdisplaybreaks                             

\usepackage{booktabs}
\usepackage{enumerate}
\usepackage{hyperref}
\hypersetup{hidelinks,
            colorlinks=true,
            allcolors=[RGB]{0,84,159}}

\usepackage[capitalise]{cleveref}               
\crefname{fig_a}{Fig.}{Fig.}                    
\crefname{fig_b}{Fig.}{Fig.}                    
\crefname{fig_c}{Fig.}{Fig.}                    
\Crefname{fig_a}{Figure}{Figure}                
\Crefname{fig_b}{Figure}{Figure}
\Crefname{fig_c}{Figure}{Figure}
\creflabelformat{fig_a}{#2{#1(a)}#3}
\creflabelformat{fig_b}{#2{#1(b)}#3}
\creflabelformat{fig_c}{#2{#1(c)}#3}

\newcommand*\subtxt[1]{_{\textnormal{#1}}}
\DeclareRobustCommand\_{\ifmmode\expandafter\subtxt\else\textunderscore\fi}

\newcommand*\supertxt[1]{^{\textnormal{#1}}}
\DeclareRobustCommand\^{\ifmmode\expandafter\supertxt\else\textasciicircum\fi}

\newcommand{\dexpval}[1]{\ensuremath{\langle \! \langle #1 \rangle \! \rangle}}

\newcommand{\vast}{\bBigg@{4}}
\newcommand{\Vast}{\bBigg@{5}}

\newcommand{\affiliationRWTH}{
Institut f\"ur Theorie der Statistischen Physik, RWTH Aachen University and JARA-Fundamentals of Future Information Technology, 52056 Aachen, Germany
}
\newcommand{\affiliationMPSD}{
Max Planck Institute for the Structure and Dynamics of Matter,
Center for Free-Electron Laser Science (CFEL),
Luruper Chaussee 149, 22761 Hamburg, Germany
}
\newcommand{\affiliationUPV}{
Nano-Bio Spectroscopy Group,
Departamento de F\'isica de Materiales,
Universidad del Pa\'is Vasco,
20018 San Sebastian, Spain
}
\newcommand{\affiliationCCQ}{
Center for Computational Quantum Physics,
Flatiron Institute, Simons Foundation,
New York City, NY 10010, USA
}

\newcommand{\affiliationNNF}{
NNF Quantum Computing Programme,
Niels Bohr Institute,
Universitetsparken 5, 2100 Copenhagen, Denmark
}

\begin{document}

\title{Cavity spectroscopy for strongly correlated systems}

\author{Lukas Grunwald}
\email{lukas.grunwald@mpsd.mpg.de}
\affiliation{\affiliationRWTH}
\affiliation{\affiliationMPSD}

\author{Emil Vi\~nas Bostr\"om}
\email{emil.bostrom@mpsd.mpg.de}
\affiliation{\affiliationMPSD}
\affiliation{\affiliationUPV}

\author{Mark Kamper Svendsen}
\affiliation{\affiliationNNF}
\affiliation{\affiliationMPSD}

\author{Dante M.~Kennes}
\affiliation{\affiliationRWTH}
\affiliation{\affiliationMPSD}

\author{Angel Rubio}
\email{angel.rubio@mpsd.mpg.de}
\affiliation{\affiliationMPSD}
\affiliation{\affiliationCCQ}
\affiliation{\affiliationUPV}

\date{\today}

\begin{abstract}
Embedding materials in optical cavities has emerged as an
intriguing perspective
for controlling quantum materials, but a key challenge lies in measuring properties of the embedded matter. Here, we propose a framework for probing strongly correlated cavity-embedded materials through direct measurements of cavity photons. We derive general relations between photon and matter observables inside the cavity, and show how these can be measured via the emitted photons. As an example, we demonstrate how the entanglement phase transition of an embedded H$_2$ molecule can be accessed by measuring the cavity photon occupation, and showcase how dynamical spin correlation functions can be accessed by measuring dynamical photon correlation functions. Our framework provides an all-optical method to measure static and dynamic properties of cavity-embedded materials.

\end{abstract}

\maketitle

\paragraph*{Introduction.--}

Controlling materials with light is a key objective of modern condensed matter research \cite{basovPropertiesDemand2017,tokuraEmergentFunctionsQuantum2017,disaEngineeringCrystalStructures2021,delatorreColloquiumNonthermal2021}. In addition to transient material modifications induced by ultrafast laser pulses, which have led to phenomena like light-induced superconductivity \cite{cavalleriPhotoinducedSuperconductivity2018,grunwaldDynamicalOnsetLightinduced2024} and quantum anomalous Hall effects \cite{mciverLightinducedAnomalous2020}, embedding matter in optical cavities has emerged as a promising approach for controlling ground state properties by coupling to quantized vacuum fluctuations of the photon field \cite{kennesNewEraQuantum2022,schlawinCavityQuantum2022,ruggenthalerUnderstandingPolaritonic2023}. Such embedding can lead to the formation of hybrid light-matter states with drastically different properties compared to free space~\cite{Hubener2024}, and can be controlled by designing the electromagnetic environment. Starting with the demonstration of the Purcell effect \cite{purcellSpontaneousEmission1964}, the growing research area of cavity material engineering now encompasses enhanced superconductivity \cite{luCavityEngineered2024,thomasExploringSuperconductivity2019}, the optical control of magnetic states \cite{Thomas2021,vinasbostromControllingMagneticState2023,weberCavityrenormalizedQuantum2023}, the breakdown and enhancement of topological protection~\cite{appuglieseBreakdownTopologicalProtection2022,Enkner2024}, and the modification of a metal-to-insulator transitions~\cite{Jarc2023}.

A challenge in this emerging field is the measurement of the embedded matter.
In general, the embedded systems have collective modes in the regime $0.25 - 2.5$ THz but dimensions on the order of $1$ $\mu$m, placing them far below the diffraction limit of standard THz spectroscopy. Recent work has tried to bridge this gap via time-resolved on-chip THz transmission measurements~\cite{Kipp2024}, or by spatially confining the probe field via THz resonators~\cite{Helmrich2024}. However, in such setups the observed material modifications might be attributed to spurious effects other than the interaction of matter with cavity vacuum fluctuations, e.g. heating due to the cavity mirrors or the dressing of probe particles~\cite{Jarc2023}. To disentangle spurious effects and the formation of strongly hybridized polaritonic states, it is hence desirable to develop a more direct probe of cavity embedded materials.
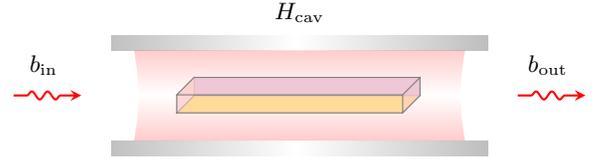
\begin{figure}
    \centering
    \newcommand{\Depth}{5}
    \newcommand{\Height}{1}
    \newcommand{\Width}{0.4}
    
    \begin{tikzpicture}[sphoton/.style={thick,->,>=stealth,decorate,decoration={snake, amplitude=1.5, segment length=6, pre length=5, post length=5}}]
    
     \begin{scope}[shift={(1.5,0)},rotate=0,scale=1.0]
     \shade[left color=lightgray, right color=lightgray, middle color=white, shading angle=90] (0,0.6) rectangle (5,0.8);
     \shade[left color=lightgray, right color=lightgray, middle color=white, shading angle=90] (0,-0.8) rectangle (5,-0.6);
     \shade[top color=red, bottom color=red, middle color=white, shading angle=0, opacity=0.2] (0.3,-0.6) -- (4.7,-0.6) to [bend left=10] (4.7,0.6) -- (0.3,0.6) to [bend left=10] (0.3,-0.6);
     \end{scope}
    
     \begin{scope}[shift={(2.6,0)},rotate=0,scale=0.6]
     \coordinate (O) at (0,0,0);
     \coordinate (A) at (0,\Width,0);
     \coordinate (B) at (0,\Width,\Height);
     \coordinate (C) at (0,0,\Height);
     \coordinate (D) at (\Depth,0,0);
     \coordinate (E) at (\Depth,\Width,0);
     \coordinate (F) at (\Depth,\Width,\Height);
     \coordinate (G) at (\Depth,0,\Height);
    
     \draw[gray,fill=yellow!80] (O) -- (C) -- (G) -- (D) -- cycle; 
     \draw[gray,fill=blue!30] (O) -- (A) -- (E) -- (D) -- cycle; 
     \draw[gray,fill=red!10] (O) -- (A) -- (B) -- (C) -- cycle; 
     \draw[gray,fill=red!20,opacity=0.8] (D) -- (E) -- (F) -- (G) -- cycle; 
     \draw[gray,fill=red!20,opacity=0.6] (C) -- (B) -- (F) -- (G) -- cycle; 
     \draw[gray,fill=red!20,opacity=0.8] (A) -- (B) -- (F) -- (E) -- cycle; 
     \end{scope}
    
     \draw[sphoton,red] (0.2, 0.0) -- node[above] {} (1.1, 0.0);
     \draw[sphoton,red] (6.9, 0.0) -- node[above] {} (7.8, 0.0);
    
     \node at ( 0.6, 0.4) {$b_{\rm in}$};
     \node at ( 7.3, 0.4) {$b_{\rm out}$};
     \node at ( 4.0, 1.1) {$H_{\rm cav}$};
    
    \end{tikzpicture}
    \caption{\textbf{Theoretical setup}. An incoming photonic wave-packet $b\_{in}(t)$ (can also be the vacuum) interacts with the system $H\_{cav}$ consisting of a material embedded in a cavity, and is scattered into a new state $b\_{out}(t)$.
    The input-output formalism provides a relation between $b\_{in}(t)$ and $b\_{out}(t)$ in terms of cavity photon correlation functions, which are in turn related to matter correlations.
    }
    \label{fig:io_sketch}
\end{figure}

A promising direction is to use the photons emitted by the cavity to extract information about the hybrid light-matter state. During its formation, the hybrid state develops finite light-matter entanglement~\cite{passettiCavityLightMatter2023,mendez-cordobaEnyiEntropySingularities2020}, such that information about matter is imprinted on the cavity photons. Recently, this connection was used to relate topological properties of matter to cavity observables, in non-interacting systems and for weak light-matter couplings~\cite{lysneQuantumOptics2023}. However, as we will demonstrate below, the relationship between light and matter observables is more general, and extends to strongly correlated systems.

In this work, we propose a scheme to probe cavity embedded correlated materials by observing cavity photons through optical measurements [\cref{fig:io_sketch}]. Demonstrating how photonic cavity observables become experimentally accessible outside the cavity through a quantum optical input-output formalism \cite{gardinerInputOutputDamped1985,drummondQuantumSqueezing2004,fanInputoutputFormalism2010}, we illustrate their utility in accessing information about the embedded matter. Specifically, we show how the entanglement phase transition of an embedded H$_2$-molecule can be detected by probing the cavity occupation. In explaining this phenomenon, we develop a versatile framework that links photonic and matter correlations functions in strongly correlated systems, where the effective low-energy theory can be obtained via a cavity Schrieffer-Wolff transformation. This framework is not restricted to static observables, but extends to dynamical correlation functions as well. We exemplify this by showing how the dynamical dimer-dimer correlation function of a small spin cluster is probed by the intensity auto-correlation function of the cavity, providing explicit analytical expressions in the limit of weak light-matter coupling. Our framework offers a new approach for probing both static and dynamic properties of cavity-embedded correlated materials.

\paragraph*{Light-matter Hamiltonian.--}
To establish the measurability of cavity correlation functions, we derive a general relation between photon correlation functions inside and outside the cavity using the quantum optical input-output formalism \cite{gardinerInputOutputDamped1985,drummondQuantumSqueezing2004,fanInputoutputFormalism2010,supplementary}. To that end, assume that our setup consists of a cavity-matter system described by the general Hamiltonian (see \cref{fig:io_sketch})
\begin{align}
    H\_{cav} &= H\_{m}({\bf a}^\dagger, {\bf a}) + \sum_n \Omega_n a_n^\dagger a_n,
    \label{eq:hamiltonian_matter}
\end{align}
with $H\_{m}({\bf a}^\dagger, {\bf a})$ denoting the light-coupled matter system and $a^{(\dagger)}_n$ the ladder operators of the $n$-th cavity mode with frequency $\Omega_n$. The system is coupled to a photonic continuum $H_{B} = \int_\nu \omega_\nu b_{\nu}^\dagger b_{\nu}$ outside the cavity via the cavity-bath coupling
\begin{align}
    H_{c} &= -i\sum_{n} \int_\nu \gamma_{n}(\nu) \left(a_n^\dagger b_\nu + b_\nu^\dagger a_n\right),
\end{align}
with ladder operators $\commutator*{b_{\nu}^\dagger}{b_{\nu'}} = \delta(\nu - \nu')$. The coupling of cavity and bath with $\gamma_n(\nu)$ models the leakage of cavity mirrors and allows photons to escape to the electromagnetic continuum. Using that the Hamiltonian is quadratic in the bath operators, they can be integrated out, and their influence on the dynamics of the cavity
solely enters via incoming $b\_{in}(t)_H$ and outgoing $b\_{out}(t)_H$ wave packet operators (in the Heisenberg picture)
\begin{align}
    b\_{in}(t)_H &= \frac{1}{\sqrt{2\pi}} \int_\nu e^{-i\nu (t - t_0)} b_\nu(t_0)_H, \\
    b\_{out}(t)\_H &= \frac{1}{\sqrt{2\pi}} \int_\nu e^{-i\nu (t - t_1)} b_\nu(t_1)_H.
    \label{eq:io_operators}
\end{align}
The reference times satisfy $t_0 < t < t_1$, such that $b^\dagger\_{in}(t)_H \ket{0}$ describes an incoming wave packet and $b^\dagger\_{out}(t)_H \ket{0}$ an outgoing one [\cref{fig:io_sketch}]. The distribution of $b\_{in}(t)_H$ specifies the initial conditions of the impinging photons, while $b\_{out}(t)_H$ characterizes the scattered state, measured after interaction with the system.

Input-output theory provides a relation between incoming and outgoing wave packet operators. For the theory to be well-behaved, the lower bound of the bath integral has to be extended to $-\infty$, viz. $\int_\nu = \int_0^\infty \dd{\nu} \approx \int_{-\infty}^{\infty} \dd{\nu}$. This assumes a cavity with high-Q factor ($\gamma \ll \Omega_n$) and employs the approximation that the dynamical modifications of the system due to the bath coupling are negligible compared to its internal dynamics~\cite{supplementary}.


Using the Markov approximation $\gamma_m(\nu) = (2\pi)^{-1/2} \sqrt{\gamma_m}$ \cite{gardinerInputOutputDamped1985,drummondQuantumSqueezing2004,fanInputoutputFormalism2010,supplementary}, the input-output relation reads
\begin{align}
    b\_{out}(t)_H &= b\_{in}(t)\_H + \sum_n \sqrt{\gamma_n} a_n(t)\_H.
    \label{eq:io_relation}
\end{align}
It directly relates the experimentally observable output operators to cavity observables. For a coherent state input $b\_{in}(t) \ket{\alpha} = \alpha(t) \ket{\alpha}$, the connected expectation values $\dexpval{AB} = \expval{(A - \expval{A}) (B - \expval{B})}$ of the output and cavity fields are directly proportional, and for a single-mode cavity
\begin{align}
    \nonumber
    &\dexpval{b\_{out}(t_1)^\dagger \dots b\_{out}^\dagger(t_m) b\_{out}(t_{m + 1}) \dots b\_{out}(t_n)} \\
    &\hspace{0.75cm}
    = \gamma^{n/2} \dexpval{a^\dagger(t_1) \dots a^\dagger(t_m) a(t_{m + 1}) \dots a(t_n)}.
    \label{eq:io_output_correlations}
\end{align}
Here we have assumed (anti-) time ordering of the (annihilation) creation operators. The relation between $\dexpval{A^n}$ and $\expval{A^n}$ is invertible and depends on $\expval{A^m}, m \leq n$ \cite{zinn-justinQuantumFieldTheory2002}, such that \cref{eq:io_output_correlations} provides a prescription for measuring cavity correlation functions by probing photons outside the cavity, e.g. achievable through homodyne detection \cite{vogelQuantumOptics2006,grynbergIntroductionQuantum2010,lysneQuantumOptics2023}.
Considering a leading order, perturbative expansion in $\gamma$ and weak input fields, the expectation values on the right-hand side of \cref{eq:io_output_correlations} can be evaluated without coupling to the bath. They hence directly probe the cavity correlations that carry information about matter degrees of freedom.

\paragraph{Entanglement transition of H$_2$.--}

To illustrate this scheme on a simple system, we consider an H$_2$ molecule embedded in a cavity. Assuming a single resonant cavity mode with linear polarization parallel to the optical dipole moment, the system is described by a single-band
Hubbard model coupled to light via the Peirls substitution~\cite{supplementary, luttingerEffectMagneticField1951}. The Hamiltonian is given by
\begin{align}
    H\_{m} = -\sum_{\expval{ij}\sigma}
       t_{ij} e^{-i \theta_{ij}} c^\dagger_{i\sigma} c_{j\sigma} - \vb{B} \cdot \vb{S}  + U \sum_i n_{i \uparrow} n_{i \downarrow},
    \label{eq:cavity_hubbard}
\end{align}
with total spin $\vb{S}$, the Peirls phase in dipole approximation given by $\theta_{ij} = (e/\hbar) (\vb{R}_j -\vb{R}_i) \cdot \hat{\vb{A}} = g_{ij} / \sqrt{N} (a^\dagger + a)$ and the cavity ladder operators satisfying $\commutator*{a^\dagger}{a} = 1$. For the H$_2$ molecule the indexes are restricted to $i, j \in \{1, 2\}$, and \cref{eq:cavity_hubbard} describes electrons hopping between two ionic cores with an on-site repulsion $U$. The Zeeman coupled magnetic field $\vb{B} = B \vb{e}_z$ tunes the ground state between an entangled spin singlet ($S = 0, B < B^*$) and a polarized spin triplet ($S = 1, B > B^*$), with electrons localized at their respective sites. We solve the system using exact diagonalization.

\begin{figure}
    \centering
    \includegraphics[width=0.9\columnwidth]{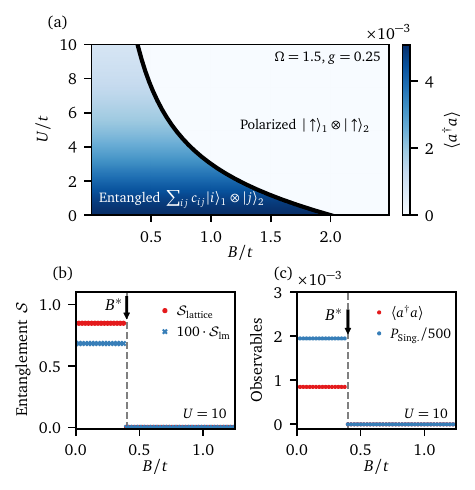}
    \caption{\textbf{Entanglement transition of the H$_2$ molecule.} (a) Entanglement phase diagram as a function of magnetic field $B$ and on site repulsion $U$. For a given $U$ there exists a critical $B^*(U)$ (black line) where the ground state transitions from an entangled spin singlet ($B < B^*$) into a polarized triplet state ($B > B^*$). The color map indicates the cavity occupation $n = \expval*{a^\dagger a}$ that acts as an order parameter for the transition. (b) Entanglement entropy as a function of magnetic field $B$ across the phase transition (gray dashed; arrow). Both dimer-dimer entanglement (red) and light matter entanglement (blue) vanish in the polarized phase. (c) Photonic occupation $\expval{a^\dagger a}$ and singlet projection operator $P\_{Sing.} = \expval{\vb{S}_1 \cdot \vb{S}_2 - 1/4}$ across the phase transition, illustrating that both provide valid order parameters.}
    \label{fig:ee_transition}
    \label[fig_a]{fig:ee_transition_a}
    \label[fig_b]{fig:ee_transition_b}
    \label[fig_c]{fig:ee_transition_c}
\end{figure}

This entanglement phase transition of the embedded molecule can be detected by observing the cavity photons, as shown in \cref{fig:ee_transition}. \cref{fig:ee_transition_a} presents the entanglement phase diagram, where the color map denotes the cavity occupation $\expval*{a^\dagger a}$. A sharp discontinuity at $B^*(U)$ (black line), marks the transition from finite to zero entanglement entropy between the lattice sites. This behavior is further emphasized in \cref{fig:ee_transition_b} (red) and \cref{fig:ee_transition_c} (red), which show the lattice site entanglement $\mathcal{S}\_{lattice}$ and cavity occupation across the phase transition for fixed $U$, respectively. Thus, the cavity occupation $\expval*{a^\dagger a}$ serves as an order parameter for the entanglement phase transition.
%

To gain deeper insights, we analyze the entanglement entropy $\mathcal{S}\_{lm}$ between light and matter across the phase transition [\cref{fig:ee_transition_b} (blue)]. It exhibits the same discontinuous behavior as $\expval*{a^\dagger a}$ and vanishes in the polarized phase, indicating that in this regime the ground state is a product state between light and matter $\ket{\textnormal{Matter}} \otimes \ket{\textnormal{Light}}$. This occurs because the polarized states span one dimensional Hilbert spaces with distinct magnetic quantum numbers $m = 0, \pm 1$, which are not connected by hopping processes and therefore cannot be dressed by the Peierls phase. As a result, the polarized state does not couple to light, explaining the discontinuous behavior of $\expval*{a^\dagger a}$ and confirming its role as an order parameter.

Including spin-orbit interactions, which manifest as spin-flip hopping processes, or multiple bands would complicate this simple argument. However, since both effects are negligible for the H$_2$ dimer, they do not invalidate the observed phenomenology. Further details, including an estimate of the magnetic field required to induce the polarization transition, are provided in the Supplement~\cite{supplementary}.


\paragraph*{General formalism.--}

These arguments offer a problem-specific explanation of how, and more importantly why, the cavity can be used to probe the entanglement structure of the embedded molecule, but
for a broad class of systems, the relationship between light and matter observables can even be made explicit through the Schrieffer-Wolff (SW) transformation~\cite{schriefferRelationAnderson1966,bravyiSchriefferWolff2011}. The SW transformation is a unitary block diagonalization procedure
\begin{align}
    \tilde{H} = e^{S} H e^{-S} = H + [S, H] + \frac{1}{2!} [S, [S, H]] + \dots,
\end{align}
that eliminates the coupling between energetically separated low- and high-energy sectors of the Hilbert space using the anti-unitary SW operator $S = -S^\dagger$. These sectors are represented by projectors $\mathcal{P}$ (low energy) and $\mathcal{Q}$ (high energy) with $\mathcal{P} + \mathcal{Q} = 1$ and $\mathcal{P} \mathcal{Q} = 0$. The new basis is defined by $\mathcal{P} \tilde{H} \mathcal{Q} = 0 = \mathcal{Q} \tilde{H} \mathcal{P}$, viz. the condition that there is no coupling between the sectors, defining an effective low-energy theory.

As a basis transformation, it does not only modify the Hamiltonian, but also leads to a corresponding dressing $A\_{eff} = e^S A e^{-S}$ of observables. Because $S$ depends on the matter degrees of freedom, this introduces an explicit dependence of the effective photonic observables on matter correlation functions through the correction introduced by $S$. To illustrate this further, we use the paradigmatic Hubbard model \cref{eq:cavity_hubbard} as a guiding example.

The SW operator $S$ can in general not be determined in closed form, and is instead computed as a perturbation expansion, in this case in $t/U \ll 1$. The coupling between low- and high-energy sectors, spanned by states with doubly occupied sites, is eliminated order-by-order. By expanding the Hamiltonian \cref{eq:cavity_hubbard} in the photon number basis, we obtain the leading order SW operator~\cite{supplementary}
\begin{align}
    \label{eq:sw_operator}
    S^{nn'} = - \Big(&
        \mathcal{P} H_t^{nn'} \mathcal{Q} \frac{1}{U + \Omega (n' - n)}
        \\
        &+
        \mathcal{Q} H_t^{nn'} \mathcal{P} \frac{1}{-U + \Omega (n' - n)}
        \Big)
        + \order*{t/U}^3,
        \nonumber
\end{align}
where $H_t^{nn'} = \mel{n}{H_t}{n'}$ is the hopping Hamiltonian evaluated between cavity number states. Using \cref{eq:sw_operator}, the effective low energy theory of \cref{eq:cavity_hubbard} takes the form of a photon dressed Heisenberg model~\cite{weberCavityrenormalizedQuantum2023,sentefQuantumClassicalCrossover2020,supplementary}
\begin{align}
    \label{eq:sw_hamiltonian}
    H\_{m,eff} &= \sum_{\expval{ij}} J_{ij}(a^{\dagger}, a)
    \left(
        \vb{S}_i \cdot \vb{S}_j - \frac{1}{4}
    \right) + \order*{t/U}^3, \\
    &\hspace{-1cm}
    J_{ij}^{nn'} = 2|t_{ij}|^2 \sum_k \phi_{ij}^{nk} \phi^{kn'}_{ji} \left( \frac{1}{U + \Omega (k - n)} + [n \leftrightarrow n'] \right)
    \nonumber
\end{align}
with the matrix element $\phi_{ij}^{nm} = \mel{n}{e^{-i \theta_{ij}}}{m}$ of the Peirls phase. Note that this expansion is only valid for $U \neq n\Omega, n \in \mathbb{Z}$, but already becomes unreliable close to these resonances due to diverging energy denominators~\cite{supplementary}.

We are in particular interested in how cavity observables are dressed in the low-energy theory. To explore this, we consider the expectation value of the cavity occupation $n = a^\dagger a$. We find~\cite{supplementary}
\begin{align}
    \expval{n\_{eff}} = \expval{n} + \sum_{\expval{ij}} \mathcal{N}_{ij} \expval{\vb{S}_i \cdot \vb{S}_j - 1/4} + \order{t/U}^4
    \label{eq:number_correction}
\end{align}
where the photon dependent prefactor is defined as $\mathcal{N}_{ij} = -\frac{U}{2} \sum_k \phi_{ij}^{0k} \phi_{ji}^{k0} k(U + \Omega k)^{-2} \in \mathbb{R}$. Note that the bare contribution gives $\expval{n} \in \mathcal{O}(t/U)^4$, since the ground state wave function has $\ket{\psi} = \ket{\psi\_m} \otimes \ket{n = 0} + \order{t/U}^2$, a universal feature for Peirls coupled tight-binding models \cite{supplementary}. \Cref{eq:number_correction} provides a direct connection between the cavity occupations and matter correlations, which is straightforward to generalize to other photon observables; see Supplement \cite{supplementary}.

For the Hubbard dimer discussed in the previous section, we have $\expval{n\_{eff}} \sim \expval{\vb{S}_1 \cdot \vb{S}_2 - 1/4}$, which corresponds to the singlet projection operator. It gives a finite value for singlet states and zero for triplet states [\cref{fig:ee_transition_c} red], offering an alternative and generalizable explanation of why the cavity occupation can be used to diagnose the entanglement phase transition discussed earlier.

\paragraph*{Measuring dynamical correlations.--}
The relationship between light and matter extends beyond static observables to dynamic correlation functions, which are routinely used to probe properties of matter \cite{zinn-justinQuantumFieldTheory2002}. Specifically, as we will show below, dynamical photon correlation functions $C(t) = \expval{\mathcal{O}_1(t) \mathcal{O}_2}$ are directly linked to dynamical matter correlators. In the case of auto-correlation functions, where $\mathcal{O}_1 = \mathcal{O}_2^\dagger$, they also provide information about the selection rule resolved excitation spectrum. This becomes evident in their frequency representation
\begin{align}
    C(\omega) = \int_0^\infty \dd{t} e^{i(\omega + i\eta)t} C(t)
    = \expval{ \mathcal{O}_1 \frac{i}{\omega + i \eta - \Delta H} \mathcal{O}_2},
    \nonumber
\end{align}
where $\Delta H = H - E\_{GS}$ is the Hamiltonian relative to the ground state energy. For auto-correlation functions, $C(\omega)$ probes the resolvent operator in state $\mathcal{O}_1 \ket{\rm{GS}}$, effectively capturing excitations generated when $\mathcal{O}_1$ acts on the ground state.

To illustrate the connection to matter and to provide a specific example for the measured correlation functions, we once again turn to the cavity embedded Hubbard model \cref{eq:cavity_hubbard}, focusing on the intensity-intensity correlation function $N_2(t) = \expval{n(t)n}$.
Down-folding to the low-energy theory followed by a perturbative expansion for weak light-matter interactions, we find that the dominant contribution probes a light-matter interaction modulated dimer-dimer correlation function of the matter \cite{supplementary}
\begin{align}
    \label{eq:N2_perturbative}
    \expval{n(t) \; n}\_{eff} &\sim
        \Bigg\langle
        \left[\sum_{\expval{i,j}} \frac{g_{ij}^2}{N} J_{ij}
        \left( \vb{S}_i \cdot \vb{S}_j - 1/4 \right) \right](t)
        \\ \nonumber
        &\hspace{0.4cm}
        \times
        \left[\sum_{\expval{k,l}} \frac{g_{kl}^2}{N} J_{kl}
        \left( \vb{S}_k\cdot \vb{S}_l - 1/4 \right) \right]
    \nonumber
    \Bigg\rangle \\ \nonumber
    &\hspace{0.4cm}
    +
    N_{2, \rm{doub.}}(t) +
    \mathcal{O}[g^6, (t/U)^4],
\end{align}
with $g_{ij} \sim \vb{A} \cdot (\vb{R}_i - \vb{R}_j)$ [see below \cref{eq:cavity_hubbard}].
The low-energy dynamics, governed by the dimer-dimer correlation function in \cref{eq:N2_perturbative}, is superimposed by the fast high-energy oscillations $N_{2, \rm{doub.}}(t)$ due to doublon-polariton excitations, which occur at a higher energy scales and are less prominent in shaping the overall behavior
\footnote{In frequency space $N_{2, \rm{doub.}}(\omega) \approx const$ for low energies.}.
Consequently, we focus on the low-energy dynamics moving forward.
%
\begin{figure}
    \includegraphics{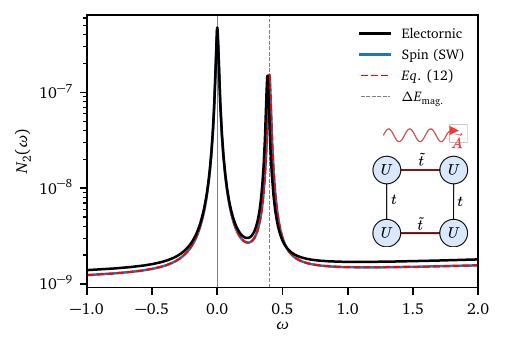}
    \caption{\textbf{Dynamical cavity correlation function $\boldsymbol{N_2(\omega)}$}
    for a $2 \times 2$ site Hubbard plaque 
    at $U = 20, \Omega = 2.35, g = 0.15$ and $\eta = 0.01$
    with an electromagentic field $\vb{A} = A \vb{e_x}$, resulting in photon dressing of hoppings in $x$ direction only. The exact electronic result (black) is compared with the approximate low energy spin theory (blue) [\cref{eq:sw_hamiltonian}], showing good agreement at low energies. The plot illustrates the zero photon peak at $\omega = 0$ and the selection rule allowed ($\Delta S, \Delta m = 0$), singlet excitation at $\omega = \Delta E\_{mag.}$ (gray dashed).
    The perturbative linear response expansion \cref{eq:N2_perturbative} (red dashed) aligns closly with the full spin result, illustrating how the photonic $N_2(\omega)$ probes the geometry modulated dimer-dimer matter correlation function.}
    \label{fig:dynamical_correlator}
\end{figure}

For uniform light-matter interactions, where $g_{ij} \equiv g$ as in 1D systems, the dimer operators in \cref{eq:N2_perturbative} commute with the Hamiltoninan leading to trivial low-energy dynamics. For inhomogeneous light matter interactions on the other hand, where $g_{ij} \neq g$, \cref{eq:N2_perturbative} probes non-trivial, dynamical dimer-dimer correlation functions. Additionally, it captures information about excitations generated by the dimer operator acting on the ground state. Since these operators commute both with $\vb{S}^2$ and $S_z$, the excitations correspond to states with $\Delta S = \Delta m = 0$.

Inhomogeneous light-matter couplings can arise from non-trivial geometries of the model. A minimal example is the $2 \times 2$ site Hubbard plaquette, where the lattice sites form a square and an electromagnetic field is applied along one of the bonds. This model geometry is depicted in the inset of \cref{fig:dynamical_correlator}, which shows $N_2(\omega)$ calculated from a full electronic model (black), a low-energy spin model (blue), and the perturbative in light-matter coupling formula (\cref{eq:N2_perturbative}, red dashed). All approaches show excellent agreement in the low-energy sector. In addition to the zero frequency peak, corresponding to no excitations, we observe a low-energy magnetic excitation at $\Delta E\_{mag}$, the only low-energy excitation of the model satisfying the desired selection rules. The excellent agreement between the full low-energy spin and perturbative approach, establishes $N_2(\omega)$ as a probe for the dimer-dimer correlation functions specified in \cref{eq:N2_perturbative}. For additional details and the analysis of alternative photonic observables like $g^{(1)}$, as well as the effect of spin-orbit interactions, see the Supplemental Material \cite{supplementary}.

\paragraph*{Conclusions.--}

We developed a framework to probe correlated, cavity-embedded materials by measuring correlation functions of the cavity photons, which we connected to matter correlation functions through a general scheme relying only on the down-folding to an effective low-energy theory. This optical method is therefore widely applicable, and enables the exploration of both static and dynamic properties of cavity-embedded materials, overcoming the diffraction limit of conventional spectroscopic techniques. Although here the cavity was used mainly as a probe of matter, an exciting possibility is to combine this framework with the study of systems where the cavity also induces a phase transition of the embedded material~\cite{luCavityEngineered2024,appuglieseBreakdownTopologicalProtection2022,Jarc2023,Latini2021,vinasbostromControllingMagneticState2023}. This allows to probe the material via correlations of the same cavity photons responsible for the modified ground state. As our framework can be extended to non-classical input states, another interesting application is the study of topological phases of matter like fractional quantum Hall states or quantum spin liquids, which are known to not be accessible with classical optical probes. Such non-classical cavity spectroscopy opens a new pathway towards probing highly entangled cavity-embedded materials via optical measurements.




\paragraph*{Acknowledgements.--}
\acknowledgments
We thank C. Le, M. Eckstein, H. Dulisch, M. Hafezi, A. Grankin and G. Nambiar for their insightful discussions on topics related to this work. We acknowledge support by the Max Planck Institute New York City Center for Non-Equilibrium Quantum Phenomena, the Cluster of Excellence 'CUI: Advanced Imaging of Matter'- EXC 2056 - project ID 390715994 and  SFB-925 "Light induced dynamics and control of correlated quantum systems" – project 170620586 of the Deutsche Forschungsgemeinschaft (DFG), and Grupos Consolidados (IT1453-22). The Flatiron Institute is a Division of the Simons Foundation.
EVB acknowledges funding from the European Union's Horizon Europe research and innovation programme under the Marie Sk{\l}odowska-Curie grant agreement No. 101106809 (CavityMag). D.M.K. acknowledge support by the DFG via Germany’s Excellence Strategy-Cluster of Excellence Matter and Light for Quantum Computing (ML4Q, Project No. EXC $2004/1$, Grant No. $390534769$) and individual grant No. 508440990. M.K.S acknowledges support from the Novo Nordisk Foundation, Grant number NNF22SA0081175, NNF Quantum Computing Programme.

\bibliographystyle{apsrev4-1}
\bibliography{bibliography.bib}

\end{document}


\title{Supplementary Material for: Cavity spectroscopy for strongly correlated systems}
\date{\today}

\author{Lukas Grunwald}
\email{lukas.grunwald@mpsd.mpg.de}
\affiliation{\affiliationRWTH}
\affiliation{\affiliationMPSD}

\author{Emil Vi\~nas Bostr\"om}
\email{emil.bostrom@mpsd.mpg.de}
\affiliation{\affiliationMPSD}
\affiliation{\affiliationUPV}

\author{Mark Kamper Svendsen}
\affiliation{\affiliationNNF}
\affiliation{\affiliationMPSD}

\author{Dante M.~Kennes}
\affiliation{\affiliationRWTH}
\affiliation{\affiliationMPSD}

\author{Angel Rubio}
\email{angel.rubio@mpsd.mpg.de}
\affiliation{\affiliationMPSD}
\affiliation{\affiliationCCQ}
\affiliation{\affiliationUPV}

\maketitle
\onecolumngrid
\tableofcontents


\section{Input-Output formalism}

We develop an Input-Output formalism \cite{gardinerInputOutputDamped1985,drummondQuantumSqueezing2004,fanInputoutputFormalism2010,wallsQuantumOptics2008}, that relates observables of the bath photons, modeled as a continuum of modes outside the cavity, to correlation functions of the cavity, in a scattering setup illustrated in \cref{fig:io_sketch}. An incoming wave-package $b\_{in}(t)$ interacts with the system and is scattered into a new state $b\_{out}(t)$. The Input-Output formalism provides a relation between $b\_{in}(t)$ and $b\_{out}(t)$ in terms of cavity correlation functions, that we derive in this section following \cite{gardinerInputOutputDamped1985,wallsQuantumOptics2008}. Additionally, we discuss a `classical derivation' in \cref{ssec:input_output_classical_analogy} following \cite{drummondQuantumSqueezing2004}, that provides further intuition for the input/output operators and the applicability of the formalism. These derivations are not novel, but translate the quantum optics literature into the more common condensed matter language.
\begin{figure}
    \centering
    \newcommand{\Depth}{5}
    \newcommand{\Height}{1}
    \newcommand{\Width}{0.4}
    
    \begin{tikzpicture}[sphoton/.style={thick,->,>=stealth,decorate,decoration={snake, amplitude=1.5, segment length=6, pre length=5, post length=5}}]
    
     \begin{scope}[shift={(1.5,0)},rotate=0,scale=1.0]
     \shade[left color=lightgray, right color=lightgray, middle color=white, shading angle=90] (0,0.6) rectangle (5,0.8);
     \shade[left color=lightgray, right color=lightgray, middle color=white, shading angle=90] (0,-0.8) rectangle (5,-0.6);
     \shade[top color=red, bottom color=red, middle color=white, shading angle=0, opacity=0.2] (0.3,-0.6) -- (4.7,-0.6) to [bend left=10] (4.7,0.6) -- (0.3,0.6) to [bend left=10] (0.3,-0.6);
     \end{scope}
    
     \begin{scope}[shift={(2.6,0)},rotate=0,scale=0.6]
     \coordinate (O) at (0,0,0);
     \coordinate (A) at (0,\Width,0);
     \coordinate (B) at (0,\Width,\Height);
     \coordinate (C) at (0,0,\Height);
     \coordinate (D) at (\Depth,0,0);
     \coordinate (E) at (\Depth,\Width,0);
     \coordinate (F) at (\Depth,\Width,\Height);
     \coordinate (G) at (\Depth,0,\Height);
    
     \draw[gray,fill=yellow!80] (O) -- (C) -- (G) -- (D) -- cycle; 
     \draw[gray,fill=blue!30] (O) -- (A) -- (E) -- (D) -- cycle; 
     \draw[gray,fill=red!10] (O) -- (A) -- (B) -- (C) -- cycle; 
     \draw[gray,fill=red!20,opacity=0.8] (D) -- (E) -- (F) -- (G) -- cycle; 
     \draw[gray,fill=red!20,opacity=0.6] (C) -- (B) -- (F) -- (G) -- cycle; 
     \draw[gray,fill=red!20,opacity=0.8] (A) -- (B) -- (F) -- (E) -- cycle; 
     \end{scope}
    
     \draw[sphoton,red] (0.2, 0.0) -- node[above] {} (1.1, 0.0);
     \draw[sphoton,red] (6.9, 0.0) -- node[above] {} (7.8, 0.0);
    
     \node at ( 0.6, 0.4) {$b_{\rm in}$};
     \node at ( 7.3, 0.4) {$b_{\rm out}$};
     \node at ( 4.0, 1.1) {$H_{\rm cav}$};
    
    \end{tikzpicture}
    \caption{\textbf{Theoretical setup}. An incoming photonic wave-packet $b\_{in}(t)$ (can also be the vacuum) interacts with the system $H\_{cav}$ consisting of a material embedded in a cavity, and is scattered into a new state $b\_{out}(t)$.
    The input-output formalism provides a relation between $b\_{in}(t)$ and $b\_{out}(t)$ in terms of cavity photon correlation functions, which are in turn related to matter correlations.
    }
    \label{fig:io_sketch}
\end{figure}

\subsection{General formalism}
\label{sec:input_output_formalism}

The full model (also defined in the main text) reads
\begin{align}
    H = H\_m(a, a^\dagger) + \sum_m \Omega_m a^\dagger_m a_m + \int_\nu \nu b_\nu^\dagger b_\nu -i \int_\nu \sum_m \gamma_m(\nu)
    \left(
        a_m^\dagger b_\nu + h.c.
    \right)
    \label{eq:io_model}
\end{align}
where $H\_m$ describes the electronic system coupled to the cavity, $a^{(\dagger)}_m$ the photons of the $m$-th cavity mode and $b_\nu^{(\dagger)}$ the photons of the continuum outside the cavity. The lower bound of the integral $\int_\nu = \int_0^\infty \dd{\nu} \eqsim \int_{-\infty}^{\infty}$ is approximated to be extended to $-\infty$. This assumes a cavity with high-Q factor, viz. $\gamma \ll \Omega_m$ so that it only interacts with frequencies $\nu \simeq \Omega_m \pm \order{\gamma}$. Additionally, it assumes a rotating wave approximation in the cavity-bath coupling, that states that the dynamical modifications due to the interaction with the bath are slow compared to the internal dynamics, viz. $\norm*{\partial_t a(t)\_I} \ll \Omega_m$, where the interaction picture is taken with respect to the cavity bath coupling. See \cref{ssec:input_output_classical_analogy} for additional discussion.

\Cref{eq:io_model} is Gaussian in the bath operators, so that the equation of motion for them can be formally solved to find
\begin{align}
    b_\nu(t)_{t'} = e^{-i\nu(t - t')} b_\nu(t')\_H + \sum_m \int^t_{t'} \dd{s} e^{-i\nu (t - s)} \gamma_m(\nu) a_m(s)\_H
    \label{eq:io_bath_evolution}
\end{align}
where we introduced the reference time $t'$, and the subscript $\textnormal{H}$ indicates that we understand the operators in the Heisenberg picture. We now define the initial conditions, viz. the \textit{Input}, as $t' = t_0 < \forall t$ and the final condition, viz. the \textit{output}, by choosing $t' = t_1 > \forall t$. Then we define the Input and Output wave-package operators
\begin{align}
    b\_{in}(t)_H &= \frac{1}{\sqrt{2\pi}} \int_\nu e^{-i\nu (t - t_0)} b_\nu(t_0)_H \qquad
    b\_{out}(t)\_H = \frac{1}{\sqrt{2\pi}} \int_\nu e^{-i\nu (t - t_1)} b_\nu(t_1)_H
    \label{eq:io_operators}
\end{align}
which behave as regular ladder operators if and only if $\int_\nu \simeq \int_{-\infty}^{\infty} \dd{\nu}$, viz.
\begin{align}
    [b\_{in}(t)_H, b\_{in}(t')_H] = \int_\nu e^{-i\nu(t - t')} = \delta(t - t').
\end{align}
From the second to last equal sign we see, that when not extending the spectrum of the harmonic oscillator to negative numbers, the commutator will involve a principle value $P(1 / (t - t'))$, stemming from the Dirac theorem. One typically sends $ t_0 \to -\infty$ and $t_1 \to \infty$, so that the wave package operators describe the asymptotic state of the bath before/after the interaction with the system. To better understand the definition \cref{eq:io_operators}, we consider the frequency representation of the input operator
\begin{align}
    b\_{in}(\omega) = \frac{1}{\sqrt{2\pi}} \int_{-\infty}^{\infty} \dd{t} e^{i \omega t} b\_{in}(t)\_H
    = e^{i\omega t_0} b_\omega(t_0)\_H
\end{align}
which is nothing but the interaction picture bath operator at $t_{0,1}$. It hence probes the occupation of bath modes at initial $t_0$ (for input) or final $t_1$ (for output) time. More concretely, we can consider the spectral occupation of the input field
$
    I(\omega) = \expval{b^\dagger\_{in}(\omega) b\_{in}(\omega)} = \expval{b_\omega^\dagger(t_0)\_H b_\omega(t_0)\_H}
 $
which measures the input spectrum/intensity at $t = t_0$. It's Fourier transform then describes (at least heuristically) the input t-domain spectrum.

The naming convention of input $b\_{in}(t)$ and output field $b\_{out}(t)$ might seem arbitrary at first glance, but it can be made more explicit by inspecting their commutation relations with an arbitrary system operator $s(t)\_H$ (cavity and/or electronic), for simplicity here with a single cavity mode. They acquire a causality structure
\cite{gardinerInputOutputDamped1985,wallsQuantumOptics2008}
\begin{align}
    [s(t)\_H, b\_{in}(t')] &= - \theta(t - t') \sqrt{\gamma} [s(t)\_H, a(t')\_H] \\
    [s(t)\_H, b\_{out}(t')] &=  \theta(t' - t) \sqrt{\gamma} [s(t)\_H, a(t')\_H]
\end{align}
viz. the input field at time $t$ only influences the system at later times, while the output field at time $t$ only depends on system degrees of freedom at earlier times.

Having gained some intuition for the wave package operators, we proceed to derive a relation between the input and output operators. For this, we only need the fact that $b_\nu(t)_{t'}$ [\cref{eq:io_bath_evolution}] is uniquely determined at time $t$. Subtracting \cref{eq:io_bath_evolution} at $t' = t_0$ and $t' = t_1$, we derive the input-output relation
\begin{align}
    b\_{out}(t)_H &= b\_{in}(t)\_H \sum_m + \int_{t_0}^{t_1} \dd{s} \Gamma_m(\nu; t - s) a_m(s)\_H,
    \label{eq:io_relation_full} \\
    \Gamma(\nu; x) &= \frac{1}{\sqrt{2\pi}} \int_\nu e^{-i \nu x} \gamma_m(\nu) \overset{\text{Markov}}{\eqsim} \delta(x) \gamma_m.
\end{align}
Since this relation is not causal with respect to $t$, but only with respect to the reference times, it is generally difficult to use in practice and one typically resorts to the \textit{Markov approximation} $\gamma_m(\nu) = (2\pi)^{-1/2} \sqrt{\gamma_m}$, which leads to the Markovian input-ouput relation
\begin{align}
    b\_{out}(t)_H &= b\_{in}(t)\_H + \sum_m \sqrt{\gamma_m} a_m(t)\_H.
    \label{eq:io_relation}
\end{align}
This provides a direct relation between input and output in terms of cavity correlation functions. This relation can in particular be used to derive explicit relations between bath and cavity correlation functions.

\subsection{Bath Correlation Functions}

In the current setup, we are interested in measuring output correlation functions of the form

\begin{align}
    \expval{:b^\dagger\_{out}(t_1) \dots b\_{out}^\dagger(t_n) b\_{out}(t_n') \dots b\_{out}(t_1'):}
    \label{eq:output_correlation}
\end{align}
where we assume wlog. that the expression is normal ordered, indicated by the double dots, and that time arguments of anihilation (creation) operators are (anti) time-ordered, viz. $t_1^{(')} < t_2^{(')} < \dots < t_n^{(')}$. Using the input-output relations \cref{eq:io_relation}, we can express \cref{eq:output_correlation} in terms of the input and system operators. The simplest relation of this type is the amplitude of the output field $E\_{out} \sim (b\_{out} + b\_{out})(t)$, which is linear in the field operators
\begin{align}
    \expval{b\_{in}(t)\_H} = \expval{b\_{out}(t)\_H} + \sqrt{\gamma} \expval{a(t)\_H}.
\end{align}
Higher order correlation functions are readily evaluated. For the 2p function (1st coherence) we find
\begin{align}
    \expval{:b^\dagger\_{out}(t)b\_{out}(t'):} = \expval{b^\dagger\_{in}(t)b\_{in}(t')} + \gamma \expval{a^\dagger(t) a(t')} +
    \sqrt{\gamma}
    \expval{b^\dagger\_{in}(t)a(t') + a^\dagger(t)b\_{in}(t')}
    \label{eq:io_2p_output}
\end{align}
The three terms have a distinct physical meaning. The expression $\order{1}$ is simply the input field, that passes through the system, while $\order{\gamma}$ measures the cavity correlation function in the presence of the input field. $\order{\sqrt \gamma}$ measures the correlations between the input field and the cavity directly. The latter terms do not factorize for a generic input signal and its determination requires the solution of the full Langevin equation \cite{gardinerInputOutputDamped1985}.

If the initial bath state is a coherent states $\bigotimes_\omega \ket{\alpha_\omega}$, all terms factorize and can be determined directly. Assuming an initial coherent state, one finds \cite{gardinerInputOutputDamped1985,wallsQuantumOptics2008}
\begin{align}
    \dexpval{:
    \tilde{\mathbb{T}}
    \big(
    b\_{out}(t_1)^\dagger \dots b\_{out}^\dagger(t_m)
    \big)
    \mathbb{T} \big(
    b\_{out}(t_{m + 1}) \dots b\_{out}(t_n)
    \big)
    :}
    = \gamma^{n/2} \dexpval{:
    \tilde{\mathbb{T}} \big(
    a^\dagger(t_1) \dots a^\dagger(t_m)
    \big)
    \mathbb{T} \big(
    a(t_{m + 1}) \dots a(t_n)
    \big):},
    \label{eq:io_output_correlations}
\end{align}
where we introduced time $\mathbb{T}$ and anti-time $\tilde{\mathbb{T}}$ ordering and the centered expectation values (\textit{fluctuation expectation values})
\begin{align}
    \dexpval{AB \dots C} = \expval{(A - \expval{A})(B - \expval{B}) \dots (C - \expval{C})}.
\end{align}
Note that generally, the relation between $\dexpval{A^n}$ and $\expval*{A^n}$ is invertible and depends on $\expval*{A^m}, \forall m \leq n$ \cite{zinn-justinQuantumFieldTheory2002}. Hence, \cref{eq:io_output_correlations} provides a full prescription to determine the coherence functions of the cavity.

\Cref{eq:io_output_correlations} involves the evaluation of the cavity correlation functions while coupling to the input field. For a high-Q cavity ($\gamma \ll g$) with weak driving fields,
we use a perturbative expansion in $\gamma$ instead. Here, the cavity expectation values can be evaluated without the coupling to the bath, viz.
$
    \dexpval{\dots} = \dexpval{\dots}_0 + \order{\gamma}.
$
Intuitively one can think of the input field amplifying the vacuum fluctuations of the cavity instead of drastically modifying the occupation of the cavity mode.

\subsection{Relation to Scattering Theory}

In this section we briefly discuss the relationship between conventional quantum mechanical scattering theory and the quantum optics Input-Output formalism, following \cite{fanInputoutputFormalism2010,xuInputoutputFormalism2015,daltonQuantumScattering1999}. Scattering theory is concerned with evaluating the scattering matrix $S$ that connects asymptotic incoming with asymptotic outgoing states \cite{taylorScatteringTheory1972}
\begin{align}
    \ket{\psi\_{out}} = S \ket{\psi\_{in}}
\end{align}
of a system interacting with an incoming pulse with Hamiltonian $H = H_0 + V$. In this case, $V$ would be the cavity bath coupling. The $S$ operator can be expressed in terms of the M\o{}ller operators $\Omega_\pm = \lim_{t \to \pm \infty} e^{i H t} e^{-i H_0 t}$ which transform free states into scattering states
\begin{align}
    \ket{\alpha}^\pm = \Omega_\pm \ket{\alpha}
\end{align}
The $S$-matrix can now be expressed in terms of the M\o{}ller operators as $S = \Omega_-^\dagger \Omega_+$, and one is typically interested in evaluating matrix elements of it, viz.
\begin{align}
    \mel*{\alpha}{S}{\beta} = \; ^-\ip*{\alpha}{\beta}^+ \equiv \mel{0}{b\_{out}(\alpha) b\_{in}^\dagger(\beta)}{0}
\end{align}
where we introduced the scattering input and output operators defined by
\begin{align}
    \label{eq:scat_input}
    b\_{in}(\alpha) &= \Omega_- b_\alpha \Omega_-^\dagger = e^{iHt_0} e^{-iH_0t_0} b_\alpha e^{iH_0t_0} e^{-iHt_0} \qquad \text{for } t_0 \to -\infty \\
    b\_{out}(\alpha) &= \Omega_+ b_\alpha \Omega_+^\dagger =
    e^{iHt_1} e^{-iH_0t_1} b_\alpha e^{iH_0t_1} e^{-iHt_1} \qquad \text{for } t_1 \to +\infty.
    \label{eq:scat_output}
\end{align}
These ladder operators create particles in the scattering states, viz. $b\_{in}(\nu) \ket{0} = \ket{\nu}^+$, where we used that $\Omega^\dagger \ket{0} = \ket{0}$, which can be argued from the form of the Hamiltonian, which always involves the absorption of particles, giving zero when acting on the vacuum. When assuming $t_{0, 1}$ large but finite (approximation), these scattering ladder operators further fullfill canonical commutation relations.

In the input-output theory of quantum optics, one defines input-output operators as given in \cref{eq:io_operators}, where $b_\nu(t_i)_H = e^{iHt_i} b_\nu e^{-iHt_i}$ in the Heisenberg picture. Using $b_\nu(t_i)_{H_0} = e^{-i\nu t_i} b_\nu$ we can rewrite these as
\begin{align}
    b\_{in}(t) = \frac{1}{\sqrt{2\pi}} \int_\nu e^{iHt_0} e^{-iH_0t_0} b_\nu e^{iH_0 t_0} e^{-i H t_0} e^{-i\nu t} = \frac{1}{\sqrt{2\pi}} \int_\nu b\_{in}(\nu) e^{-i \nu t}
\end{align}
where we used the definition of the input scattering operator \cref{eq:scat_input}. In the limit $t_0 \to -\infty$ and $t_1 \to \infty$, the quantum optical input output operators, hence correspond to the Fourier transform of the scattering input-output operators, that create particles in the scattering states $\ket{\nu}^+$. Both formalisms, can hence be used to evaluate $S$-Matrix elements \cite{fanInputoutputFormalism2010,xuInputoutputFormalism2015} and correlation functions. They are hence equivalent in that regard. The evaluation of output field correlation functions is not equivalent to evaluating $S$ matrix elements though, but provides a complementary approach.

\subsection{Classical Quantization -- Model and Setup}
\label{ssec:input_output_classical_analogy}

\begin{figure}[h]
    \centering
    \includegraphics[width = 0.5\linewidth]{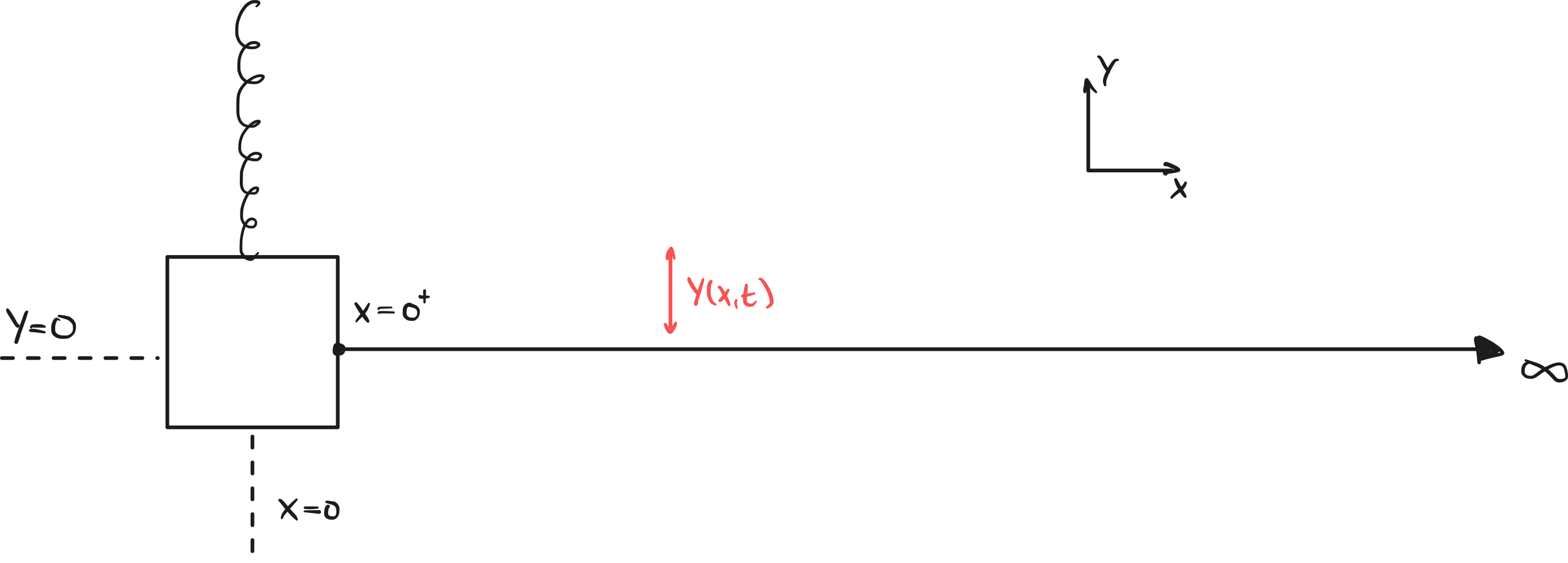}
    \caption{\textbf{Classical analogy for Input-Output formalism}. A harmonic oscillator at $x = 0$ interacts with a string at $x > 0$. The string acts as a thermal bath and a driving force for the harmonic oscillator and resembles the vacuum field outside the cavity in the setup discussed in the main text.}
    \label{fig:io_sketch_classical}
\end{figure}

In this section we discuss the input-output formalism in the context of a canonically quantized classical theory, following \cite{drummondQuantumSqueezing2004}. Our classical model, illustrated in \cref{fig:io_sketch_classical} consists of a one dimensional harmonic oscillator (cavity) at $x = 0$, coupled to a string (electromagnetic vacuum) at $x > 0$. The string acts as the thermal bath and driving force for the HO, providing a situation in close analogy the setup discussed in the conventional derivation \cref{sec:input_output_formalism}.

Since transmission line and harmonic oscillator are connected at $x = 0$, we can use the same variable $Y(x, t)$ to describe the displacement of both. The motion of the HO is then modeled by $Y(x = 0, t)$, while the displacement of the string is described by $Y(x > 0, t)$. The system is modeled by the classical Lagrangian
\begin{align}
    \mathcal{L}(x, t) = \delta(x) \left\{
        \frac{M}{2} \qty( \pdv{Y}{t} )^2 - V(Y)
    \right\} +
    \theta(x) \left\{
        \frac{\rho}{2} \qty( \pdv{Y}{t} )^2 - \frac{\sigma}{2} \qty( \pdv{Y}{x} )^2
    \right\}.
\end{align}
Here $M$ is the mass of the oscillator and $V(Y) = k \; Y(x = 0, t)^2$ the harmonic oscillator potential. The string is characterized by string density $\rho$ and string tension $\sigma$. Using the
Euler-Lagrange equations, we derive a coupled system of PDE's for the displacement of oscillator and string
\begin{align}
    M \dv[2]{Y}{t} + \pdv{V}{Y} = \sigma \pdv{Y}{x}\bigg\vert_{x = 0} \quad &\text{for } x = 0, \\
    \qty( \pdv[2]{Y}{t} ) - v^2 \qty( \pdv[2]{Y}{x} ) = 0 \quad &\text{for } x > 0,
    \label{eq:appdx_io_eq_motion}
\end{align}
where we introduced the abbreviation $\dv[n]{Y}{t} = \pdv[n]{Y}{t}\big\vert_{x = 0}$ and $v^2 = \sigma / \rho$. We observe that dynamical equation for the string has the general form of a wave equation, where the coupled oscillator provides a Von-Neumann boundary condition at $x = 0$. The general solution of this wave equation can be given in terms of its Fourier modes as a superposition
\begin{align}
    Y(x, t) = \frac{1}{\sqrt{4\pi \Gamma}} \int_0^{\infty} \frac{\dd{\omega}}{\sqrt{\omega}}
    \left(
        a\_L(\omega) e^{-i\omega(t + x / v)} + a\_R(\omega) e^{-i\omega(t - x / v)}
        + h.c.
    \right)
    \label{eq:appdx_io_general_wave_solution}
\end{align}
with suggestively named constants $a\_L(\omega)$ for left moving and $a\_R(\omega)$ for right moving excitations and the impedance $\Gamma = \sqrt{\sigma \rho}$. The prefactors in \cref{eq:appdx_io_general_wave_solution} are chosen such that if $Y(x, t)$ and $\Pi(x, t) = \partial \mathcal{L} / \partial (\partial_t Y)$ full-fill the canonical commutation relations upon canonical quantization, that is
\begin{align}
    \comm{Y(x, t)_H}{\Pi(x', t)_H} = i \delta(x - x'), \quad
    \comm{Y(x, t)_H}{Y(x', t)_H} = \comm{\Pi(x, t)_H}{\Pi(x', t)_H} = 0,
\end{align}
the associated ladder operators will full-fill
\begin{align}
    \comm{a\_{L/R}(\omega)}{a^\dagger\_{L/R}(\omega)} = \delta(\omega - \omega'),
    \quad
    \comm{a^{(\dagger)}\_{L/R}(\omega)}{a^{(\dagger)}\_{L/R}(\omega)} = 0,
\end{align}
viz. they can be interpreted as conventional ladder operators creating/annihilating left or right moving particles. In the following we are dealing with a fully quantum mechanical, canonically quantized, problem with all momentum and position operators full-filling the canonical commutation relations.

Looking back at \cref{eq:appdx_io_general_wave_solution}, we observe the general structure of ingoing and outgoing wave amplitude operators
\begin{align}
    Y(x, t) = Y\_{in}(t + x / V) + Y\_{out}(t - x / v),
    \label{eq:appdx_io_input_output}
\end{align}
viz. one part the propagates into the harmonic oscillator, the input $Y\_{in}$, and one contribution moving away from it, the output $Y\_{out}$. Evaluating \cref{eq:appdx_io_input_output} at $x = 0$, we directly obtain an input-output relation for the harmonic oscillator at $x = 0$, viz.
\begin{align}
    Y(t) = Y\_{in}(t) + Y\_{out}(t).
\end{align}
Since the output is moving away form the system, it will not interact with the HO. To make this explicit, we can rewrite the dynamical equation of the HO in terms of $Y\_{in}$ alone. To that end, note that
$
    \pdv{Y}{x} = \frac{2}{v} \pdv{Y\_{in}}{t} - \frac{1}{v} \pdv{Y}{t}
$
which follows from combining $\pdv{Y}{t}$ and $\pdv{Y}{x}$ evaluated from \cref{eq:appdx_io_input_output}. Inserting this into \cref{eq:appdx_io_eq_motion} we arrive at a dynamical equation for the system displacement, coupled to the input field alone
\begin{align}
    M \dv[2]{Y}{t} + \Gamma \dv{\Gamma}{t} + \pdv{V}{Y} = 2\Gamma \dv{Y\_{in}}{t}.
    \label{eq:appdx_io_dynamical_damping}
\end{align}
The string hence provides a radiative damping for dynamics via $\dv{\Gamma}{t}$, while $\dv{Y\_{in}}{t}$ describes the driving force that the incoming field excerpts on the oscillator. Up to this point, the treatment did not involve any approximations, but in the following we will approximate \cref{eq:appdx_io_dynamical_damping}, to arrive at equations resembling the ones of the input-output formalism.

Before proceeding, we comment on the time dependence of the operators after the canonical quantization. When canonically quantizing $Y(x, t)$, we implicitly do this for the isolated system, i.e. with Hamiltonian
\begin{align}
    H_0 = \int_0^\infty \dd{\omega} \left( a\_L^\dagger a\_L + a\_R^\dagger a\_R \right)(\omega),
\end{align}
leading to the explicit time dependence
\begin{align}
    Y(x, t)_{H_0} \sim
    a\_L(\omega) e^{-i\omega(t + x / v)} + a\_R(\omega) e^{-i\omega(t - x / v)}
    = a\_L(\omega)_{H_0} e^{-i\omega x / v} + a\_R(\omega)_{H_0} e^{+i\omega x / v}
\end{align}
If $H_0$ is only part of the full Hamiltonian, we can write a similar expression with the operators in the interaction picture with respect to $H_0$
\begin{align}
    Y(x, t)_{H} \sim a\_L(\omega)\_I e^{-i\omega(t + x / v)} + a\_R(\omega)\_I e^{-i\omega(t - x / v)}
\end{align}
where we defined $A(t)_H = U\_I^\dagger(t, t_0) A(t_0)_{H_0} U\_I(t, t_0) = [A(t_0)_{H_0}]\_I$ with $U\_I$ the regular picture propagator. In this case the interaction picture is defined with respect to the bath cavity coupling.

\subsection{Classical Quantization -- Approximations towards Input-Output formalism}

For concreteness, we consider $V(Y) = K/2 Y^2$ in the following, i.e. a harmonic oscillator without non-linearity, which would correspond to an empty cavity in \cref{sec:input_output_formalism}. The results of the last section can then be summarized in the dynamical equation for the harmonic oscillator
\begin{align}
   &\dv[2]{Y}{t} + \gamma \dv{\Gamma}{t} + \omega_0^2 Y(t) = 2\gamma \dv{Y\_{in}}{t}, \\
   & Y\_{in}(t)_H = \frac{1}{\sqrt{4 \pi \Gamma}} \int_0^\infty \frac{1}{\sqrt \omega} \left(
    a\_{in}(\omega)_I e^{-i\omega t} + a^\dagger\_{in}(\omega)_I e^{i\omega t}
   \right)
\end{align}
where all operators are to be understood in the Heisenberg/interaction picture. We introduced the damping $\gamma = \Gamma / M$ and eigenfrequency of the harmonic oscillator $\omega_0^2 = K / M$. The interaction picture in the 2nd row follows from the discussion at the end of the last section.

Let us first decompose $Y(t)$ into ladder operators of the Harmonic oscillator
\begin{align}
    Y(t)_H = \frac{1}{\sqrt{2M\omega_0}} \left(a(t)_H + a(t)^\dagger_H \right) =
    \frac{1}{\sqrt{2M\omega_0}}\left(a(t)\_I e^{-i \omega_0 t} + a(t)^\dagger\_I e^{-i \omega_0 t} \right).
    \label{eq:appdx_io_ho_ladder_operators}
\end{align}
When the damping of the oscillator is small $\gamma \ll \omega_0$ (the quality factor of the cavity is large), the oscillator only responds to frequencies very close to its resonance frequency. In other words, the response function of the oscillator only has a finite weight for $\omega_0 \pm \order{\gamma}$, viz. $\chi(\omega) \sim \delta_\gamma(\omega - \omega_0)$. Using this insight, we can approximate the input field as
\begin{align}
    Y\_{in}(t)\_H &= \frac{1}{\sqrt{4\pi \Gamma}}
    \left\{
    e^{-i\omega_0 t} \int_{-\omega_0}^\infty \frac{\dd{\omega}}{\sqrt \omega} a\_{in}(\omega + \omega_0)(t)\_I e^{-i\omega t}
    + e^{i\omega_0 t} \int_{-\omega_0}^\infty \frac{\dd{\omega}}{\sqrt \omega} a^\dagger\_{in}(\omega + \omega_0)(t)\_I e^{i\omega t}
    \right\} \\
    &\approx
    \frac{1}{\sqrt{4\pi \Gamma \omega_0}} \int_{-\infty}^{\infty} \dd{\omega}
    \bigg(
        e^{-i \omega_0 t} a\_{in}(\omega + \omega_0)(t)\_I e^{-i \omega t}   +
        e^{i \omega_0 t} a^\dagger\_{in}(\omega + \omega_0)(t)\_I e^{i \omega t}
    \bigg) + "\order{\gamma / \omega_0}" ,
    \label{eq:appdx_io_input_approx_intermediate}
\end{align}
where we approximated $\sqrt{\omega + \omega_0} \approx \sqrt{\omega_0}$ and $\int_{-\omega_0}^{\infty} \approx \int_{-\infty}^\infty$, using that integrand `is only relevant' for $\omega_0 \pm \order{\gamma}$.
Further, we assumed that the dynamical change of the cavity ladder operators due to the interaction with the environment is much smaller than the internal dynamics, viz. $\norm*{\partial^{(n)} a_{\textnormal{in}, \omega + \omega_0}(t)\_I} \ll \omega_0$, so that the interaction picture time dependence of the ladder operators can be neglected for this argument.

Renaming $b_{\textnormal{in}, \omega} = a\_{in}(\omega + \omega_0)$, we can define input operators, (very similar to the input-output formalism) by collecting all modes in a \textit{wave package operator}
\begin{align}
    b\_{in}(t)\_H = \frac{1}{\sqrt{2 \pi}} \int_{-\infty}^{\infty} \dd{\omega} b_{\textnormal{in}, \omega}(t)\_H e^{-i \omega t}.
\end{align}
Referencing \cref{eq:appdx_io_general_wave_solution}, this makes the interpretation of the input operators as creating wave packages from the main text explicit. Using the same approximations as in \cref{eq:appdx_io_input_approx_intermediate}, one can show that $b\_{in}(t)\_H = e^{-i \omega_0 t} b\_{in}(t)\_I$, as well as that the operators full-fill the commutation relations
\begin{align}
    \comm{b\_{in}(t)}{b^\dagger\_{in}(t')} = \delta(t - t'), \quad \comm{b^{(\dagger)}\_{in}(t)}{b^{(\dagger)}\_{in}(t')} = 0.
\end{align}
With these assumptions, the input field can hence be written as
\begin{align}
    Y\_{in}(t)\_H = \frac{1}{\sqrt{4\pi \Gamma \omega_0}}
    \left(
        e^{- i \omega_0 t} b\_{in}(t)\_I + e^{ i \omega_0 t} b\_{in}(t)\_I
    \right).
    \label{eq:appdx_io_input_approximation}
\end{align}
Inserting the ladder operator representations \cref{eq:appdx_io_ho_ladder_operators} and \cref{eq:appdx_io_input_approximation} into the dynamical equation \cref{eq:appdx_io_eq_motion}, we obtain
\begin{align}
    { \color{red} \ddot{a}(t)\_I } -2i \omega_0 \left( \dot{a}(t)\_I + \frac{\gamma}{2} a(t)\_I \right) =
    2\sqrt{\gamma} \left( {\color{red} \dot{b}\_{in}(t)\_I} - i\omega_0 b\_{in}(t)\_I \right),
\end{align}
where we will neglect the terms marked in red against $\omega_0$ in the following. This assumes that the dynamical changes due to the system bath-coupling are much smaller than the internal system dynamics, viz. $\norm*{\partial_t^{(n)} a(t)\_I} \ll \omega_0$ and $\norm*{\partial_t^{(n)} b\_{in}(t)\_I} \ll \omega_0$.
This is the so called \textit{Rotating Wave Approximation} (RWA), which in this case leads to
\begin{align}
    \dv{~}{t} a(t)\_H + i \omega_0 a(t)\_H  + \frac{\gamma}{2} a(t)\_H = \sqrt{\gamma} b\_{in}(t)\_H
\end{align}
where we have transformed the equation back from the interaction to the Heisenberg picture. Apart from an arbitrary sign in the definition of the input operators, this is exactly the same dynamical equation we find when analyzing the Heisenberg equations of motion for \cref{eq:io_model} (without matter). Inserting the representation of the amplitudes $Y(x, t)$ and $Y\_{in}(t, x)$ in terms of ladder operators in \cref{eq:appdx_io_input_output} we also recover the input-output relation from the main text
\begin{align}
    b\_{out}(t)\_H = \sqrt{\gamma} a(t)\_H - b\_{in}(t)\_H,
\end{align}
completing our derivation. We have hence successfully recovered the equations from the conventional derivation of the input-output formalism. The analogy between the two derivation is not perfect, but it provides urgently needed intuition for the range of validity and the approximations needed in the input-output formalism.

Summarizing, we made two assumptions in the derivation of equations resembling the input-output formalism from the main text:
\begin{enumerate}[(i)]
    \item Weak system-bath coupling $\gamma \ll \omega_0$
    \item Rotating wave approximation $\norm*{\partial_t^{(n)} a(t)\_I} \ll \omega_0$ and $\norm*{\partial_t^{(n)} b\_{in}(t)\_I} \ll \omega_0$
\end{enumerate}
Assumption $(i)$ entails that the cavity only responds to frequencies close to its resonance frequency, which allowed us to extend the integration range $\int_{-\omega_0}^{\infty} \eqsim \int_{-\infty}^\infty$ and approximate the frequency dependence of the integrand's, since only $a\_{in}(\omega) \sim \delta_\gamma(\omega - \omega_0)$ contributes significantly. The rotating wave approximation (RWA) in the system environment coupling, viz. assumption (ii) assumes that the dynamical modifications due to the cavity-bath coupling are small compared to $\omega_0$. Intuitively, this is reasonable, since the `bath' is big and slow compared to cavity system.

\section{Hubbard Dimer and the H\texorpdfstring{$_2$}{2} Molecule}

\subsection{Ab-Initio Modeling}

In this section we determine microscopic parameters for the (extended) Hubbard model from a single band tight-binding (TB) approach, following \cite{kadzielawaH2H22014,spalekOptimizationSingleparticleBasis2000}, with the purpose to find an estimate for the magnetic field necessary to induce the entanglement transition discussed in the main text. The H$_2$ molecule consists of two protons and two electrons and, within the Born-Oppenheimer approximation, is modeled by the Hamiltonian
\begin{align}
    H = h_1 + h_2 + v(\vb{x}_1 - \vb{x_2}), \quad \text{with } h_i = \frac{\vb{p}_i^2}{2m} - v(\vb{x}_i - \vb{R}_1) - v(\vb{x}_i - \vb{R}_2),
\end{align}
where $v(\vb{x}) = e^2 / |\vb{x}|$ is the Coulumb-Potential and $\vb{R}_i$ denotes the classical position of the $i$-th proton.
We solve the system using a low-energy tight-binding approach, including only the 1$s$ atomic-orbitals $\psi_{1s}(\abs{\vb{r}}) = 1/\sqrt{\pi \alpha^3} e^{-r/\alpha}$ of the Hydrogen atom, motivated by the energy separation of $\sim 10eV$ to the next orbital (2s). The parameter $\alpha$ controls the spread of the wave function and will be subsequently optimized to minimize the ground state energy. The atomic orbitals on the different sites
\begin{align}
    \psi_1(\vb{r}) = \psi_{1s}(\vb{r} - \vb{R}_1), \quad
    \psi_2(\vb{r}) = \psi_{1s}(\vb{r} - \vb{R}_2)
\end{align}
need to be orthogonalized using a Gram-Schmidt procedure, since $\ip{\psi_1}{\psi_2} \neq 0$ in general. The orthogonalization yields Wannier-orbitals $\varphi_i$, which we use to express the general second quantized two-site Hamiltonian
\begin{align}
    \label{eq:h2_molecule}
    H &= \sum_{ij}\sum_{\sigma} t_{ij} c^\dagger_{i\sigma} c_{j \sigma}
    + \frac{1}{2} \sum_{ijkl} \sum_{\sigma \sigma'}
    v_{ijkl} c^\dagger_{i \sigma} c^\dagger_{j \sigma'} c_{l \sigma'} c_{k \sigma}
    \\ \nonumber
    &=
    \epsilon \sum_i n_i +
    t \sum_{\sigma}(c_{1\sigma}^\dagger c_{2\sigma} + h.c.) +
    U \sum_i n_{i \sigma} n_{i \bar{\sigma}}
    + \left(V\_c - \frac{V\_x}{2} \right) n_1 n_2 \\ \nonumber
    &\hspace{0.5cm}- 2V\_x \vb{S}_1 \cdot \vb{S}_2 + V\_x
    \left( c^\dagger_{1\uparrow} c^\dagger_{1\downarrow}c_{2\downarrow}c_{2\uparrow} + h.c. \right)
    + V\_{cx} \sum_\sigma (n_{1\bar{\sigma}} + n_{2\bar{\sigma}})(c^\dagger_{1\sigma} c_{2\sigma} + h.c.)
\end{align}
with coupling constants defined as (using the shorthand $\ket{\varphi_i} = \ket{i}$)
\begin{gather}
    \epsilon = \mel{1}{h}{1}, \quad
    t = \mel{1}{h}{2}, \quad
    U = \mel{1 1}{v}{1 1} \\
    V\_c = \mel{1 2}{v}{1 2}, \quad
    V\_x = \mel{1 1}{v}{2 2}, \quad
    V\_{cx} = \mel{1 1}{v}{1 2}.
\end{gather}
They are functions of the inter-atomic distance $\vb{R} = \vb{R}_1 - \vb{R}_2$ and the wave function spread $\alpha$ and their functional dependence can be determined analytically by solving the corresponding integrals \cite{kadzielawaH2H22014}. The equilibrium distance $\vb{R}$ and wave function spread $\alpha$ are then determined by optimizing the ground state energy of \cref{eq:h2_molecule}, which we find via exact diagonalization of the half-filled system. The ground state energy of the electronic Hamiltonian is supplemented with the contribution from the ion Coulomb-repulsion $1 / R$, viz.
\begin{equation}
    E(R, \alpha) = E\_{GS}(H[R, \alpha]) + e^2 / R,
\end{equation}
and subsequently optimized in $\alpha$ for every $R$. The ground state $E_-$ is a site-entangled spin singlet with an equilibrium inter-ionic distance of $R_0 \approx 1.43 a_0$ and binding energy $E\_b = 4.02eV$, which agrees well with experimental values. The corresponding equilibrium values of the microscopic coupling parameters are shown \cref{tab:h2_couplings}. Note that while the exact parameters differ significantly from a pure Hubbard model, the general phenomenology of the H$_2$ ground state is well captured by the simple Hubbard-dimer, used in the main text.
\begin{table}[h!]
    \centering
    \begin{tabular}{lc}\toprule
        \multicolumn{2}{c}{Microscopic Parameters $[E_h]$}\\
        \midrule
        $\epsilon$ & -0.875 \\
        $t$ &        -0.364 \\
        $U$ &         0.827 \\
        $V\_{c}$ &    0.478 \\
        $V\_{x}$ &    0.011 \\
        $V\_{cx}$ &  -0.0059 \\
        \bottomrule
    \end{tabular}
    \caption{Microscopic coupling parameters of single band tight-binding model for H$_2$ molecule, at the equilibrium configuration $R_0 \approx 1.43 a_0$ and $\alpha = 0.838$, in agreement with \cite{kadzielawaH2H22014}.}
    \label{tab:h2_couplings}
\end{table}

\begin{SCfigure}[][h]
    \centering
    \includegraphics{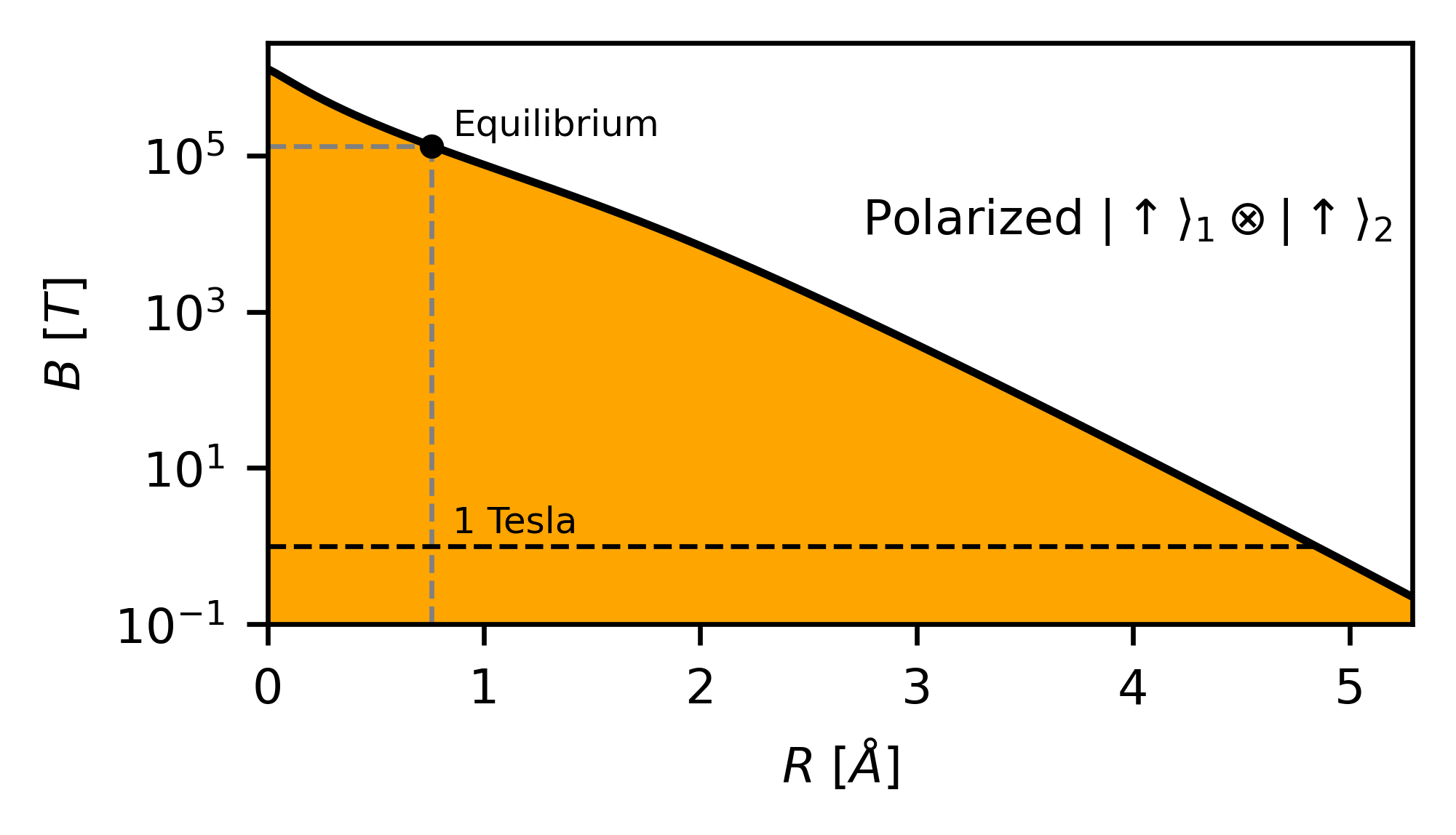}
    \caption{\textbf{Critical magnetic field to induce singlet $\boldsymbol{\to}$ triplet transition} in equilibrium, as a function of the interionic distance. The B-field necessary to induce the transition at equilibrium (grey lines) is experimentally inaccessible. When increasing the interionic distance it quickly reaches experimentally accessible regimes (black line). This can e.g. be achieved by disassociating the H$_2$ molecule (see main text). }
    \label{fig:h2_magentic field}
\end{SCfigure}

We are ultimately interested in studying a transition from entangeled singlet to polarized triplet, as it could be induced by a Zeeman coupling
\begin{align}
    H\_Z = -\boldsymbol{\mu} \cdot \vb{B} = g\_s \mu\_B \vb{S} \cdot \vb{B}
\end{align}
where $\mu\_B \approx 5.79 ~ 10^{-5} J/T$ is the Bohr magneton and $g\_s \approx 2$ is the Lande $g$-factor. This quantum phase transition appears when the energy of singlet and triplet state coincide. Explicitly, this happens at a critical magnetic field
\begin{equation}
    B_* = - \frac{U - V\_c}{2} - 2V\_x + \frac{1}{2} \sqrt{16(t + V\_{cx})^2 + (U - V\_c)^2}.
\end{equation}
At equilibrium $R = R_0$, the critical magnetic field $B_* = 0.56 E_h \approx 132.5 kT$ is experimentally inaccessible by orders of magnitude. But, when thinking about disassociating the H$_2$ molecule (H$_2 \to $ 2H), i.e. effectively increasing $\vb{R}$, the necessary magnetic field quickly reaches experimentally accessible regimes. The functional dependence of $B_*$ on the interionic distance $R$ is shown in \cref{fig:h2_magentic field}.

In terms of experimental protocols, we imagine an external pulse exciting the system from singlet into the three times degenerate triplet state with $\Delta E = 15eV$. The decay from triplet to singlet state should then be very slow, since it can only happen via processes involving spin-orbit coupling. Once in the triplet state, the system will naturally relax to larger interionic distances, making it easily polarizable, allowing for a realization of the here proposed protocol.

\subsection{Extensions of the model}

In the main text we observed that the cavity occupation $\expval*{a^\dagger a}$ provides an order parameter for the entanglement phase transition due to vanishing light matter interactions in the polarized phase. This was due to vanishing hopping processes between the polarized states with distinct magnetic quantum numbers $ m = 0, \pm 1$, which span one dimensional subspaces respectively. This restriction can be lifted by two extensions, that we want to discuss in this section.

First the inclusion of multiple bands in the Hubbard dimer. There light could couple to on-site inter-orbital dipol moments in the polarized phase. This type of coupling is neglected in the Peirls substitution \cite{luttingerEffectMagneticField1951}, but could be added by hand. Since the inter-orbital dipol moment is distinctly different to the inter-site dipol moment, we would still observe a significant jump in the cavity occupation as we cross the polarization phase boundary. Hence, the phenomenology described in the main text, where the cavity occupation can be used to diagnose the phase transition still holds.

\begin{figure}[b]
    \centering
    \includegraphics{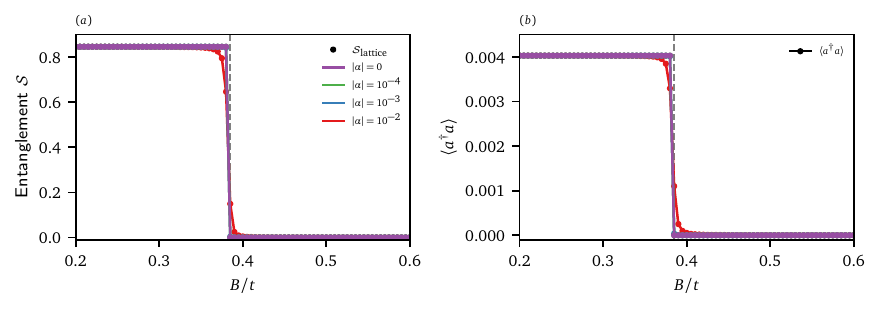}
    \caption{\textbf{Dimer entanglement transition with finite $\boldsymbol{L \cdot S}$ interactions} for $U = 10, \Omega = 0.5, g = 0.5$ and various uniform spin orbit interactions $\boldsymbol{\alpha} = \alpha e^{i \pi / 4} \mathbb{I}_3$ [see \cref{eq:appdx_sw_hubbard} for full Hamiltoninan]. (a) Lattice entanglement entropy across the phase transition, with colors denoting different strengths of spin-orbit interaction in comparison with (b) the color matched cavity occupation. We observe excellent agreement and preservation of the sharp transition for weak spin-orbit strength.}
    \label{fig:appdx_eetransition_ls}
\end{figure}

Secondly, one can include spin-orbit interactions ($\vb{L} \cdot \vb{S}$), modeled as spin dependent hopping, viz. $\sum_{\expval{i,j}} c^\dagger_{i, \cdot} (\boldsymbol{\alpha}_{ij} \cdot \boldsymbol{\sigma}) c_{j, \cdot}$ [see \cref{eq:appdx_sw_hubbard} for full Hamiltoninan]. But since these contributions are very small in the H$_2$ molecule, we still expect a jump in the cavity occupation upon crossing into the polarized phase. Strictly speaking, the spin quantum numbers are no longer good quantum numbers in this case, and there is no polarization transition as described before. But we can still compare the `jump' in lattice entanglement transition and cavity occupation, which show good agreement illustrated in \cref{fig:appdx_eetransition_ls}. Hence, the cavity occupation can still be used to diagnose the phase transition.

\section{Schrieffer-Wolff transformation of the Hubbard Model}
\label{sec:appdx_sw}

Here we provide the general derivation of the Schrieffer-Wolff transformation of the Hubbard model, emphasizing the modifications due to light-matter interactions.

\subsection{General Formalism}
The Schrieffer-Wolff transformation (SW) \cite{schriefferRelationAnderson1966,bravyiSchriefferWolff2011} is a unitary transformation that block diagonalizes the Hamiltonian to eliminate the coupling between energetically separated subspaces of the Hilbert space, in the following described by projectors $\mathcal{P}$ (low energy) and $\mathcal{Q}$ (high energy), with $\mathcal{P} + \mathcal{Q} = 1$ and $\mathcal{P} \mathcal{Q} = 0$. It hence describes a basis transformation, that acts as
\begin{align}
    H \ket{\psi} = E \ket{\psi}
    \qquad \leftrightarrow& \qquad
    \tilde{H} \ket{\phi} = E \ket{\phi}
    \quad \text{with} \quad
    \tilde{H} = e^S H e^{-S},
    \ket{\phi} = e^{S} \ket{\phi} \\
    \ev{A}{\psi}
    \qquad \leftrightarrow& \qquad
    \ev{\tilde A}{\phi}
    \quad \text{with} \quad
    \tilde{A} = e^{S} A e^{-S}
\end{align}
with the anit-unitary SW operator $S^\dagger = -S$, which usually has to be determined perturbatively, in the case of the Hubbard model as a power series in $t / U$. The new basis is then defined by demanding $\mathcal{P} \tilde{H} \mathcal{Q} = 0$, viz. that high and low energy sector are decoupled in the rotated basis. For notational convenience we define projections of the Hamiltonian that act on the different energy sectors as
\begin{align}
    H_0 = \mathcal{P} H \mathcal{P}, \quad
    H_1 = \mathcal{Q} H \mathcal{Q}, \quad
    H_2 = \mathcal{P} H \mathcal{Q} + \mathcal{Q} H \mathcal{P}.
\end{align}
To obtain a perturbative solution for the SW, expand the SW transformation in powers of $S$ as a series of commutators
\begin{align}
    \tilde{H} = e^{S} H e^{-S} = H + \sum_{i = 0}^\infty \frac{1}{n!} [S, [S, \dots [S, H]] \dots ]_n
    = H + \commutator{S}{H} + \frac{1}{2!} \comm*{S}{\commutator{S}{H}} + \order*{S}^3.
    \label{eq:general_sw_transformation}
\end{align}
The defining equation for the $S$ operator is then obtained, by demanding that $\mathcal{P} \tilde{H} \mathcal{Q} = 0$ up to the desired order in $S$, which in the case of a 2nd order expansion leads to
\begin{align}
    H_2 + \commutator{S}{H_0 + H_1} = 0, \quad \mathcal{P} S \mathcal{Q} = 0.
    \label{eq:general_2nd_order_S}
\end{align}
The associated projected low energy Hamiltonian is given by
\begin{align}
    H\_{eff} \coloneqq
    \mathcal{P} \tilde{H} \mathcal{P} = H_0 +
    \frac{1}{2}
    \left(
    \mathcal{P} S \mathcal{Q} H \mathcal{P} - \mathcal{P} H \mathcal{Q} S \mathcal{P}
    \right) + \order{S}^3
    \label{eq:general_effective_hamiltonian}
\end{align}

Formally, we can solve \cref{eq:general_2nd_order_S}, by expanding it in the exact eigenstates of $H_0 + H_1$, which can be split as $\{ \ket{x} \} = \{ \ket{x} \in \mathcal{P}\} \cup \{ \ket{x} \in \mathcal{Q} \}$, viz. high and low energy eigenstates. After a short calculation we find
\begin{align}
    S = \sum_{ml} \frac{H_{2, ml}}{E_m - E_l} \dyad{m}{l}, \quad \text{with} \quad E_x = \ev{H_0 + H_1}{x}
    \label{eq:general_formal_sw}
\end{align}
Note that this expression only gives a finite contribution if $\ket{m} \in \mathcal{Q} \leftrightarrow \ket{l} \in \mathcal{P}$ and vice versa, as demanded by \cref{eq:general_2nd_order_S}. Note that this expression is only valid for $E_m \neq E_l$, but since $\ket{m}, \ket{l}$ stem from different energy sectors.

While the formal solution of the Schrieffer-Wolff (SW) transformation is useful for the numerical implementation, it can be immensely useful to derive analytical expressions for $S$ that do not involve the exact eigenstates. The Feshbach projection, which provides an alternative to obtain an effective low-energy Hamiltonian on $\mathcal{P}$ opens a way to obtain an analytical form of the $S$ operator by comparing coefficients. To that end, rewrite the Schrödinger equation as
\begin{gather}
    H \ket{\psi} = E \ket{\psi} =
    (\mathcal{P} + \mathcal{Q}) H (\mathcal{P} + \mathcal{Q}) =
    E (\mathcal{P} + \mathcal{Q}) \ket{\psi}
    \label{eq:feshbach_1}
    \\ \to
    \left( \mathcal{P} H \mathcal{P} + \mathcal{P} H \mathcal{Q} \right) \ket{\psi} = E \mathcal{P} \ket{\psi}
    \label{eq:feshbach_2}
    \\ \to
    \left( \mathcal{Q} H \mathcal{Q} + \mathcal{Q} H \mathcal{P} \right) \ket{\psi} = E \mathcal{Q} \ket{\psi}.
    \label{eq:feshbach_3}
\end{gather}
Solving \cref{eq:feshbach_3} for $\mathcal{Q} \ket{\psi}$ and inserting into \cref{eq:feshbach_2}, we obtain an effective low energy equation
\begin{align}
    \left(
    \mathcal{P} H \mathcal{P} + \mathcal{P} H \mathcal{Q} \frac{1}{E - \mathcal{Q} H \mathcal{Q}} \mathcal{Q} H \mathcal{P}
    \right) \ket*{\tilde \psi } = E \ket*{\tilde \psi}, \quad \text{with} \quad
    \ket*{\tilde \psi} = \mathcal{P} \ket{\psi}
    \label{eq:feshbach_final}
\end{align}
In many practical cases, one can approximate $E$ appearing on left-hand side perturbatively, so that \cref{eq:feshbach_final} resembles \cref{eq:general_effective_hamiltonian}, allowing the identification of the $S$ operator.

\subsection{The Hubbard Model}
\label{sec:hubbard}

We proceed by explicitly deriving the SW transform for the electronic Hubbard model at half filling $\mu = 0$. Here we include $\vb{L} \cdot \vb{S}$-interactions for completeness. The model Hamiltonian reads

\begin{align}
    H = \sum_{\expval{i,j}} \sum_{\sigma \sigma'}
    M^{\sigma \sigma'}_{ij} c^\dagger_{i, \sigma} c_{j, \sigma'} + U \sum_i n_{i \uparrow} n_{i \downarrow} - B S\^z = H_t + H_U + H_B,
    \label{eq:appdx_sw_hubbard}
\end{align}
with $\vb{M}_{ij} = -t_{ij} \vb{1} + \boldsymbol{\alpha}_{ij} \cdot \boldsymbol{\sigma} $, the first term describing the hopping, while the 2nd term models spin-orbit interactions. We also included a Zeeman coupled magnetic field $B$, that can be used to induce a polarization phase transition.

The goal of the SW transformation is to eliminate the doubly occupied states (doublons) ($E\_{doublon} \sim U$), and hence obtain an effective theory without charge excitations, viz. where the electrons are restricted to their respective lattice site. The SW transform achieves this decoupling of doublons in a perturbative strong coupling expansion in $\order{t/U}$. The different Hamiltonian contribution introduced in the last section read
\begin{align}
    H_0 &= \mathcal{P} H \mathcal{P} = \mathcal{P} H_B \mathcal{P} \\
    H_1 &= \mathcal{Q} H \mathcal{Q} = H_U + \mathcal{Q}H_B\mathcal{Q} \\
    H_2 &= \mathcal{Q} H \mathcal{P} + \mathcal{P} H \mathcal{Q} = H_t
\end{align}
where $\mathcal{P}$ projects onto singly occupied sites, while $\mathcal{Q}$ selects states with a finite number of doublons. In the last line we used that $\mathcal{Q} H_t \mathcal{Q} = 0$, since a hopping process either gives zero, or produces a state with a doubly occupied site. The low energy strong coupling Hamiltoninan is then obtained by solving
\begin{align}
    H\_{eff} = &H_B + \frac{1}{2}
    \left(
    \mathcal{P} S \mathcal{Q} H_t \mathcal{P} - \mathcal{P} H_t \mathcal{Q} S \mathcal{P}
    \right) + \order{t/U}^3 \\
    &H_t + [S, H_U + H_B] = 0, \quad \mathcal{P} S \mathcal{Q} = 0.
\end{align}
The SW operator can be identified from the associated Feshbach projection [\cref{eq:feshbach_final}]
\begin{align}
    H\_{eff, Fesh.} =
    \left(
        \mathcal{P}H_B\mathcal{P} + \mathcal{P} H_t \mathcal{Q} \frac{1}{E - H_U - H_B} \mathcal{Q} H_t \mathcal{P}
    \right)
\end{align}
where we can replace $H_U = U$ in the denominator, since the hopping produces exactly one doublon (or gives 0). The associated energy $E$ can be approximated perturbatively as
\begin{align}
    E = 0 - B m + \order*{t/U}^2
\end{align}
with $m$ the magnetic quantum number. Without spin-orbit interactions, the magnetic contribution in the denominator is canceled exactly, since $\mathcal{Q} H_t \mathcal{P}$ only has processes with $\Delta m = 0$, so that $H_B = -B m$. With spin-flip hopping terms, this is no longer the case and the hopping can change the magnetic quantum number $m$, leading to a more complicated form. With spin-orbit interaction, the following derivation is valid for $B = 0$. The effective Hamiltonian, obtained from the Feshbach projection reads
\begin{align}
    H\_{eff, Fesh.} = \mathcal{P} H_B \mathcal{P} - \frac{1}{U} \mathcal{P} H_t \mathcal{Q} H_t \mathcal{P} + \order*{t/U}^3,
    \label{eq:hubbard_feshbach}
\end{align}
so that we can read of the SW operator by comparing with \cref{eq:general_effective_hamiltonian} to find
\begin{align}
    S = -\frac{1}{U}
    \left( \mathcal{P} H_t \mathcal{Q} - \mathcal{Q} H_t \mathcal{P} \right) + \order*{t/U}^3.
    \label{eq:hubbard_sw_operator}
\end{align}
This expression can be used to derive an explicit spin-model for the low energy sector. This is the topic of the next section.

\subsection{Effective Spin Hamiltonian}

Inspecting \cref{eq:hubbard_feshbach}, we observe that the calculation of the effective low-energy spin model involves the evaluation of $\mathcal{P} H_t \mathcal{Q} H_t \mathcal{P}$. Since we are exactly at half-filling, the projected hopping always creates a doublon or gives 0, so that the $\mathcal{Q}$ acts as a unity and can be dropped. The calculation hence amounts to evaluating $\mathcal{P} H_t^2 \mathcal{P}$, which only gives a finite contribution if the indices are matched, because else we retain a state with a finite number of doublons.
\begin{align}
    \mathcal{P} H_t^2 \mathcal{P} &=
    \mathcal{P}
    \left(
        \sum_{\expval{i,j}}\sum_{\sigma_1 \sigma_2} M^{\sigma_1 \sigma_2}_{ij} c^\dagger_{i \sigma_1} c_{j \sigma_2}
    \right)
    \left(
        \sum_{\expval{i',j'}}\sum_{\sigma_1' \sigma_2'} M^{\sigma_1' \sigma_2'}_{i'j'} c^\dagger_{i' \sigma_1'} c_{j' \sigma_2'}
    \right) \mathcal{P}
    \\&=
    \mathcal{P}
    \sum_{\expval{a, b}_{\textnormal{pairs}}}
    \sum_{\sigma_1 \dots \sigma_4}
    \left\{
        M^{\sigma_1 \sigma_2}_{ab} M^{\sigma_3 \sigma_4}_{ba}
        c^\dagger_{a \sigma_1} c_{a \sigma_4}
        \left( \delta_{\sigma_2 \sigma_3}
             - c^\dagger_{b \sigma_3} c_{b \sigma_2} \right)
        + (a \leftrightarrow b)
    \right\} \mathcal{P}
\end{align}
Using the identity $c^\dagger_{i \sigma} c_{i \sigma'} = \left( \vb{S}_i \cdot \boldsymbol{\sigma} + n_i / 2 \right)_{\sigma' \sigma}$ and the fact that we are at half-filling, such that $n_i \equiv 1$, the final Hamiltonian can be written as a trace over the spin structure
\begin{align}
    H\_{eff} = H_B
    - \frac{1}{U} \sum_{\expval{i, j}} \sum_{\sigma_1 \dots \sigma_4}
    \Bigg\{
    \Tr[
        \vb{M}_{ij} \vb{M}_{ji} \big( \vb{S}_i \cdot \boldsymbol{\sigma} + 1 / 2 \big)
    ]
    -
    \Tr[
        \big( \vb{S}_i \cdot \boldsymbol{\sigma} + 1 / 2 \big) \vb{M}_{ij}
        \big( \vb{S}_j \cdot \boldsymbol{\sigma} + 1 / 2 \big) \vb{M}_{ji}
    ]
    + (i \leftrightarrow j)
    \Bigg\}
\end{align}
where we dropped the projection operators for convenience. Using computatinoal relations for Pauli matrices $\Tr(\boldsymbol{\sigma}) = 0$ and $(\vb{a} \cdot \boldsymbol{\sigma}) (\vb{b} \cdot \boldsymbol{\sigma}) = (\vb{a} \cdot \vb{b}) + i (\vb{a} \cross \vb{b}) \cdot \boldsymbol{\sigma}$, this expression can be simplified. One finds the effective spin model
\begin{align}
    \nonumber
    H\_{eff} &= \sum_{\expval{i,j}}
    \left(
        \sum_{\mu \nu} S_i^\mu \Delta^{\mu \nu}_{ij} S_j^\nu + \xi_{ij}
    \right) - B \sum_i S_i\^z + \order*{t/U}^3
    \\
    &=
    \sum_{\expval{i, j}}
    \Bigg\{
        J_{ij} \bigl( \vb{S}_i \cdot \vb{S}_j - 1/4 \bigr) -  \Lambda_{ij} \bigl( \vb{S}_i \cdot \vb{S}_j + 1/4 \bigr)
        + \vb{D}_{ij} \cdot \bigl( \vb{S}_i \cross \vb{S}_j \bigr)
        + S_i^\mu \Gamma_{ij}^{\mu \nu} S_j^\nu
    \Bigg\} - B \sum_i S_i\^z + \order*{t/U}^3
    \label{eq:hubbard_spin_model}
\end{align}
where the vector, scalar and tensor quantities correspond to (corrected) exchange, Dzyaloshinskii-Moriya and symmetric anisotropy interactions, which are given by
\begin{align}
    \label{eq:hubbard_spin_constants_1}
    \Delta_{ij}^{\mu \nu} &=
    \frac{4}{U} \left( |t_{ij}|^2 - |\boldsymbol{\alpha}_{ij}|^2 \right) \delta^{\mu \nu} +
    \frac{8}{U} \epsilon_k^{\; \mu \nu }
    \textnormal{Im}(t_{ij}\boldsymbol{\alpha}_{ji})^k
    + \frac{8}{U} \textnormal{Re}(\alpha_{ij}^\mu \alpha_{ji}^\nu)
    \equiv
    \bigl( J_{ij} - \Lambda_{ij} \bigr) \delta^{\mu \nu} + \epsilon_k^{\mu \nu} (\vb{D}_{ij})^k + \Gamma_{ij}^{\mu \nu},
    \\
    \xi_{ij} &= -\frac{1}{U} \left(  |t_{ij}|^2 + |\boldsymbol{\alpha}_{ij}|^2 \right).
    \label{eq:hubbard_spin_constants_2}
\end{align}
Here we used the symmetry $t_{ij}^\dagger = t_{ji}, \boldsymbol{\alpha}_{ij}^\dagger = \boldsymbol{\alpha}_{ji}$ and the fact that both index pairs $(i, j)$ and $(j, i)$ give the same contribution. We verified the validity of \cref{eq:hubbard_spin_model} by analyzing the error scaling of the eigenenergies of spin and exact electronic model.

\subsection{Hubbard Model in a Cavity}
\label{sec:hubbard_cavity}

We proceed to derive an effective low energy spin theory for the Hubbard model embedded into a cavity, which is coupled to the tight-binding model via a Peirls substitution \cite{luttingerEffectMagneticField1951}. Assuming a single relevant effective cavity mode at frequency $\Omega$, the Hamiltonian reads
\begin{align}
    H = \sum_{\expval{i,j}} \sum_{\sigma \sigma'}
    \left(
    M^{\sigma \sigma'}_{ij} e^{-i \theta_{ij}} c^\dagger_{i, \sigma} c_{j, \sigma'}  + h.c. \right)
    + U \sum_i n_{i \uparrow} n_{i \downarrow} - B S\^z + \Omega \left(a^\dagger a + \frac{1}{2} \right)
    \equiv H_t + H_U + H_B + H_\Omega
    \label{eq:cavity_hamiltonian}
\end{align}
with the Peirls phase in the dipol approximation $\theta_{ij} = e \int_{\vb{R}_j}^{\vb{R}_i} \dd{\vb{x}} \cdot \vb{A}(\vb{x}, t) \approx g_{ij} / \sqrt{N} (a^\dagger + a)$ and the cavity ladder operators $\commutator*{a^\dagger}{a} = 1$. For one dimensional models we typically have $g_{ij} \equiv g$.

While it is possible to derive the SW transformation, while keeping the photonic operator structure \cite{sentefQuantumClassicalCrossover2020}, a more convenient formulation for us can be found by expanding the Hamiltonian \cref{eq:cavity_hamiltonian} in the photonic number operator basis $\sum_n \dyad{n}{n} = \vb{I}$, yielding
\begin{align}
    H &= \sum_{nm} H_{nm} \dyad*{n}{m} \\
    H_{nm} &= \sum_{\expval{i, j}} \left( c_{i, .}^\dagger \vb{M}^{nm}_{ij} c_{j, .} + h.c. \right) + \delta_{nm}
    \left[
        H\_U + H\_B + \Omega \left(n  + \frac{1}{2} \right)
    \right]
\end{align}
where we introduced the matrix elements $\vb{M}^{nm}_{ij} = \vb{M}_{ij} \mel{n}{e^{-i \theta_{ij}[a, a^\dagger]}}{m} \equiv \vb{M}_{ij} \phi_{ij}^{nm} $. Note in particular the factorization into photonic and electonic contribution of the matrix elements that we will use explicitly below.

As before, the SW transform decouples the high energy doublon states $\mathcal{Q} \ket{\psi}$ from the low energy spin theory $\mathcal{P} \ket{\psi}$ perturbatively in $\order*{t/U}$. Hence, the projection operators $\mathcal{Q}, \mathcal{P}$ commute with the photonic operators, resulting in the factorization
\begin{align}
    H_0 &= \mathcal{P} H \mathcal{P} = \mathcal{P} (H_B + H_\Omega) \mathcal{P} \\
    H_1 &= \mathcal{Q} H \mathcal{Q} = H_U + \mathcal{Q} (H_B + H_\Omega) \mathcal{Q} \\
    H_2 &= \mathcal{Q} H \mathcal{P} + \mathcal{P} H \mathcal{Q} = H_t
\end{align}
where we used $\mathcal{P} H_t \mathcal{P} = 0$ in the last line. We determine the SW operator using the formal solution of the SW equation
\begin{align}
    H\_{eff} = &H_B + H_\Omega + \frac{1}{2}
    \left(
    \mathcal{P} S \mathcal{Q} H_t \mathcal{P} - \mathcal{P} H_t \mathcal{Q} S \mathcal{P}
    \right) + \order{t/U}^3
    \label{eq:cavity_Heff_intermediate}
    \\
    &H_t + [S, H_U + H_B + H_\Omega] = 0, \quad \mathcal{P} S \mathcal{Q} = 0,
\end{align}
by expanding in the exact eigenstates of $H_0 + H_1$, which are product states of light and matter, $\ket*{l, n} = \ket{l} \otimes \ket{n}$, since the LM coupling only enters in $H_2$. Using \cref{eq:general_formal_sw} and setting $B = 0$ (see \cref{sec:hubbard}), we can write the matrix elements as
\begin{align}
    S^{n n'}_{l l'} = - \frac{\mel{l}{H_t^{nn'}}{l'}}{E_{l'}^{n'} - E_l^n}
    = - \mel{l}{H_t^{nn'}}{l'}
    \left(
        \frac{\delta_{l \in \mathcal{P}, l \in \mathcal{Q}}}{U + \Omega(n - n')} +
        \frac{\delta_{l \in \mathcal{Q}, l \in \mathcal{P}}}{-U + \Omega(n - n')}
    \right),
\end{align}
or more conveniently, using that $\sum_{ll'} \dyad{l}{l'} = \vb{I}$, written with the explicit electronic operators
\begin{align}
    S^{nn'} = - \left(
        \mathcal{P} H\_t^{nn'} \mathcal{Q} \frac{1}{U + \Omega (n' - n)} +
        \mathcal{Q} H\_t^{nn'} \mathcal{P} \frac{1}{-U + \Omega (n' - n)}
    \right) + \order*{t/U}^3.
    \label{eq:cavity_sw_operator}
\end{align}
Inserting \cref{eq:cavity_sw_operator} into \cref{eq:cavity_Heff_intermediate} and using that $H_t$ produces exactly one doublon (or gives 0), like in the electronic model, the effective Hamiltonian becomes
\begin{align}
    H\_{eff} = H_\Omega - \frac{1}{2} \sum_k \mathcal{P} H_t^{nk} H_t^{kn'} \mathcal{P}
    \left(
        \frac{1}{U + \Omega(k - n)} + \frac{1}{U + \Omega(k - n')}
    \right)
\end{align}
In an analog procedure to the electronic model, we hence have to derive an explicit expression for $ \mathcal{P} H_t^{nk} H_t^{kn'} \mathcal{P}$, which can be written as a trace operation over the spin structure
\begin{align}
    \nonumber
    \mathcal{P}H_t^{nk} H_t^{kn'}\mathcal{P} = \sum_{\expval{i,j}}
     &\Bigg\{
        \frac{1}{2} \Tr[
            \vb{M}_{ij} \vb{M}_{ji}( \vb{S}_i \cdot \boldsymbol{\sigma})
        ] (\phi_{ij}^{nk}\phi_{ji}^{kn'} - \phi_{ji}^{nk}\phi_{ij}^{kn'})
        \\
        &-
        \left(
        \Tr[
            (\vb{S}_i \cdot \boldsymbol{\sigma}) \vb{M}_{ij}
            (\vb{S}_j \cdot \boldsymbol{\sigma}) \vb{M}_{ji}
        ]
        - \frac{1}{4} \Tr[\vb{M}_{ij} \vb{M}_{ji}]
        \right) \phi_{ij}^{nk} \phi_{ji}^{kn'}
        + (i \leftrightarrow j)
        \Bigg\},
\end{align}
where we have separated out the photonic contribution in terms of the Peirls factors $\phi_{ij}^{nm} = \mel{n}{e^{-i \theta_{ij}[a, a^\dagger]}}{m}$. This can be evaluated using the Pauli matrix identities, yielding an explicit expression for the spin-cavity system. One finds for the effective Hamiltonian
\begin{align}
    H\_{eff} = \sum_{\expval{i,j}}
    \left\{
    \left(
        \sum_{\mu \nu} S_i^\mu \Delta^{\mu \nu}_{ij} S_j^\nu + \xi_{ij}
    \right) \mathcal{D}_{ij}^+
    + \left(
        (\vb{S}_i - \vb{S}_j) \cdot \boldsymbol{\zeta}\^s_{ij} -i (\vb{S}_i + \vb{S}_j) \cdot \boldsymbol{\zeta}_{ij}\^a
    \right) \mathcal{D}_{ij}^-
    \right\}
    + \Omega \left(a^\dagger a + \frac{1}{2}\right),
    \label{eq:cavity_Heff}
\end{align}
where we introduced the photonic dressing operators $\mathcal{D}^\pm_{ij}$, given by
\begin{align}
    \mathcal{D}^\pm_{ij} = \frac{U}{4}
    \sum_{nn'}\sum_k
    \left( \phi_{ij}^{nk} \phi_{ji}^{kn'} \pm \phi_{ji}^{nk} \phi_{ij}^{kn'} \right)
    \left( \frac{1}{U + \Omega (k - n)} + \frac{1}{U + \Omega (k - n')} \right) \dyad*{n}{n'},
    \label{eq:cavity_photon_dressing}
\end{align}
and identified constants $\Delta^{\mu \nu}_{ij}$ and $\xi_{ij}$ which are exactly the same as in the electronic model [viz. \cref{eq:hubbard_spin_constants_1} and \cref{eq:hubbard_spin_constants_2}]. The constants $\boldsymbol{\zeta}\^{s/a}$ are novel contributions that only appear in the presence of the cavity, which breaks a detailed balance between hopping processes leading to their cancelation in the matter system. They are given by
\begin{align}
    \boldsymbol{\zeta}_{ij}\^s = \frac{4}{U} \textnormal{Re}(t_{ij}\boldsymbol{\alpha}_{ji}), \quad
    \boldsymbol{\zeta}_{ij}\^a = \frac{2}{U} \boldsymbol{\alpha}_{ij} \cross \boldsymbol{\alpha}_{ji}
\end{align}
and in particular enter the Hamiltonian with a different photonic prefactor $\sim \mathcal{D}^-_{ij}$, as compared to the spin-interactions $\sim \mathcal{D}^+$. These linear terms vanish without spin-orbit interactions, leading to the Hamiltonian reported in the main text. The validity of the low energy model \cref{eq:cavity_Heff} can again be assessed by considering the error scaling of the eigen-energies of spin and full electronic model.

\subsection{Understanding the resonances of $\mathcal{D}^\pm_{ij}$}
\label{ssec:hubbard_cavity_resonance}

The perturbative expansion in $\order{t/U}$ is formally valid for $U \neq n\Omega, n \in \mathbb{Z}$, since else the energies in the formal solution of the SW operator [\cref{eq:general_formal_sw}] become degenerate and the expression invalid. An intuitive picture emerges, when noting that the resonance condition above states that photons can resonantly create charge excitations in the doublon sector, so that the energetic seperation between the half-filled and doubly occupied sector breaks down. Practically, the perturbative SW transformation already becomes unreliable close to these resonances because the neglected correction terms will also involve terms similar to $\mathcal{D}^\pm$ and hence become large due to the vanishing denominators \cref{eq:cavity_photon_dressing}.

\begin{figure}[t]
    \centering
    \includegraphics{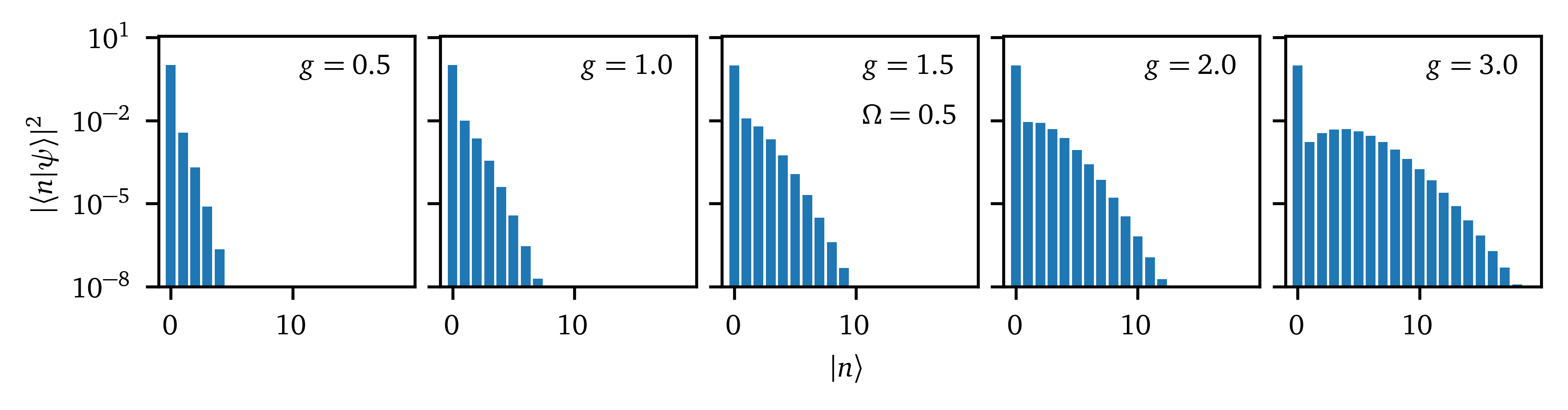}
    \caption{\textbf{Ground state photon occupation distribution} for $U/t = 10$ and $\Omega/t = 0.5$ without spin-orbit coupling at various $g$ with $n\_c = 120$ the bosonic cutoff. The majority of weight resides in the $\ket{n = 0}$ state, while the occupation in all higher order number states is at least $10^2$ times smaller. For larger $g$, more $n > 0$ states become important, but the separation of scales for $n = 0$ and $n > 0$ is universal.}
    \label{fig:sw_photon_occupation}
\end{figure}

The light-matter interaction strength $g$ controls the width of the resonances. To understand this we analyze the overlap $\abs{(\bra{n} \otimes \mathbb{1}) \ket{\psi}}^2$ with $\ket{\psi}$ the electronic ground state, which measures the spectral weight in the different $\ket{n}$ states. \Cref{fig:sw_photon_occupation}, shows the result of this analysis and we observe that for larger $g$, larger $\ket{n}$ become relevant for the ground state, giving the corresponding terms in \cref{eq:cavity_photon_dressing} a significant weight. The resonance hence contributes further away from divergence of the term, spoiling the approximation. We typically consider the limit $\Omega \ll U$, at weak light matter interactions $g \approx 0.15$, where a tiny shift away from the resonances produces accurate results.

\section{Observables in the strong coupling expansion}

The Schrieffer-Wolff transformation does not only change the Hamiltonian, but also leads to a dressing of the observables in exactly the same manner
\begin{align}
    A\_{eff} = A + [S, A] + \frac{1}{2}[S, [S, A]] + \order*{t/U}^3,
    \label{eq:sw_operator_transformation}
\end{align}
Calculating the effective photonic observables for the SW transformed model, reveal corrections that are dependent on electronic observables, hence providing an explicit connection between matter and light. In the following we will analyze this further. The main object entering the evaluation of \cref{eq:sw_operator_transformation} is the product of two SW operators with varying photonic indices
\begin{align}
    \label{eq:sw_ss_product}
    S^{nk} S^{ln'} &= \mathcal{P} H_t^{nk} H_t^{{ln'}} \mathcal{P} \frac{1}{[U + \Omega (k - n)][U + \Omega(l - n')]} + \mathcal{Q} \dots \mathcal{Q}
    \\
    &=
    \frac{U}{2} \sum_{\expval{i, j}}
    \bigg\{
    \left( \vb{S}_i\^T \boldsymbol{\Delta}_{ij} \vb{S}_j + \xi_{ij} \right)
    \Phi^{+, nkln'}_{ij}
    + \left(
    \vb{S}_i \cdot \boldsymbol{\zeta}^-_{ij}
    + \vb{S}_j \cdot \boldsymbol{\zeta}^+_{ij}
    \right)
    \Phi^{-, nkln'}_{ij}
    \bigg\}
    \frac{1}{[U + \Omega (k - n)][U + \Omega(l - n')]} + \mathcal{Q} \dots \mathcal{Q}
    \nonumber
    \\
    &\equiv
    \frac{U}{2} \sum_{\expval{i, j}}
    \bigg\{
    (\vb{SS})^+_{ij} \Phi^{+, nkln'}_{ij} + \vb{S}^-_{ij} \Phi^{-, nkln'}_{ij}
    \bigg\}
    \frac{1}{[U + \Omega (k - n)][U + \Omega(l - n')]} + \mathcal{Q} \dots \mathcal{Q}
\end{align}
where we defined $\boldsymbol{\zeta}^\pm_{ij} = \boldsymbol{\zeta}\^s_{ij} \pm i \boldsymbol{\zeta}\^a_{ij}$ and the shorthand $\Phi^{\pm, nkln'}_{ij} = \phi_{ij}^{nk} \phi_{ji}^{ln'} \pm \phi_{ji}^{nk} \phi_{ij}^{ln'}$ as well as the abbreviations for the quadratic $(\vb{SS})^+_{ij}$ and linear $\vb{S}^-_{ij}$ spin operators. Note that for vanishing spin-orbit interactions $\vb{S}^-_{ij} \equiv 0$. The contribution $\mathcal{Q} \dots \mathcal{Q}$ is not evaluated explicitly, since it only contributes at higher orders, then considered in our perturbative expansion. Armed with this relation, the evaluation of \cref{eq:sw_operator_transformation} amounts to inserting \cref{eq:sw_ss_product} and collecting terms. In the next section we consider single operators, in particular the number $n$ and ladder $a^{(\dagger)}$ operators.

\subsection{Scalar Observables}

\subsubsection*{Number Operator}

We begin this section, by deriving an explicit expression for the number operator. Using \cref{eq:sw_operator_transformation} and \cref{eq:sw_ss_product}, one finds
\begin{align}
    \nonumber
    n\_{eff} &= n + \sum_{nn'} S^{nn'}(n' - n)  \otimes  \dyad*{n}{n'} +
    \mathcal{P} \sum_{\expval{i,j}}
    \bigg\{
        (\vb{SS})^+_{ij} \otimes \mathcal{N}^+_{ij} + \vb{S}^-_{ij} \otimes \mathcal{N}^-_{ij}
    \bigg\} \mathcal{P}
    + \mathcal{Q} \dots \mathcal{Q}
    + \order*{t/U}^3 \\
    &= n + \delta n + \sum_{nn'} S^{nn'}(n' - n)  \otimes  \dyad*{n}{n'} + \order*{t/U}^3,
    \label{eq:sw_number_occupation}
\end{align}
with the photonic correction operators
\begin{align}
    \mathcal{N}^\pm_{ij} = \frac{U}{2} \sum_{nn'} \sum_k
    \left(
        \phi_{ij}^{nk} \phi_{ji}^{kn'} \pm \phi_{ji}^{nk} \phi_{ij}^{kn'}
    \right) \frac{\frac{1}{2}(n + n') - k}{[U + \Omega (k - n)] [U + \Omega (k - n')]}
    \dyad*{n}{n'},
    \label{eq:sw_number_photon_correction}
\end{align}
and the shorthand $\delta n = \sum_{\expval{i,j}} \left\{ (\vb{SS})^+_{ij} \otimes \mathcal{N}^+_{ij} + \vb{S}^-_{ij} \otimes \mathcal{N}^-_{ij} \right\}$. Note that since the number operator is diagonal in the number-basis \cref{eq:sw_number_photon_correction} remains valid for any function of the number operator $f(n)$ with the replacement $\frac{1}{2}(n + n') - k \to \frac{1}{2}(f(n) + f(n')) - f(k)$.

\begin{SCfigure}[][h]
    \centering
    \includegraphics{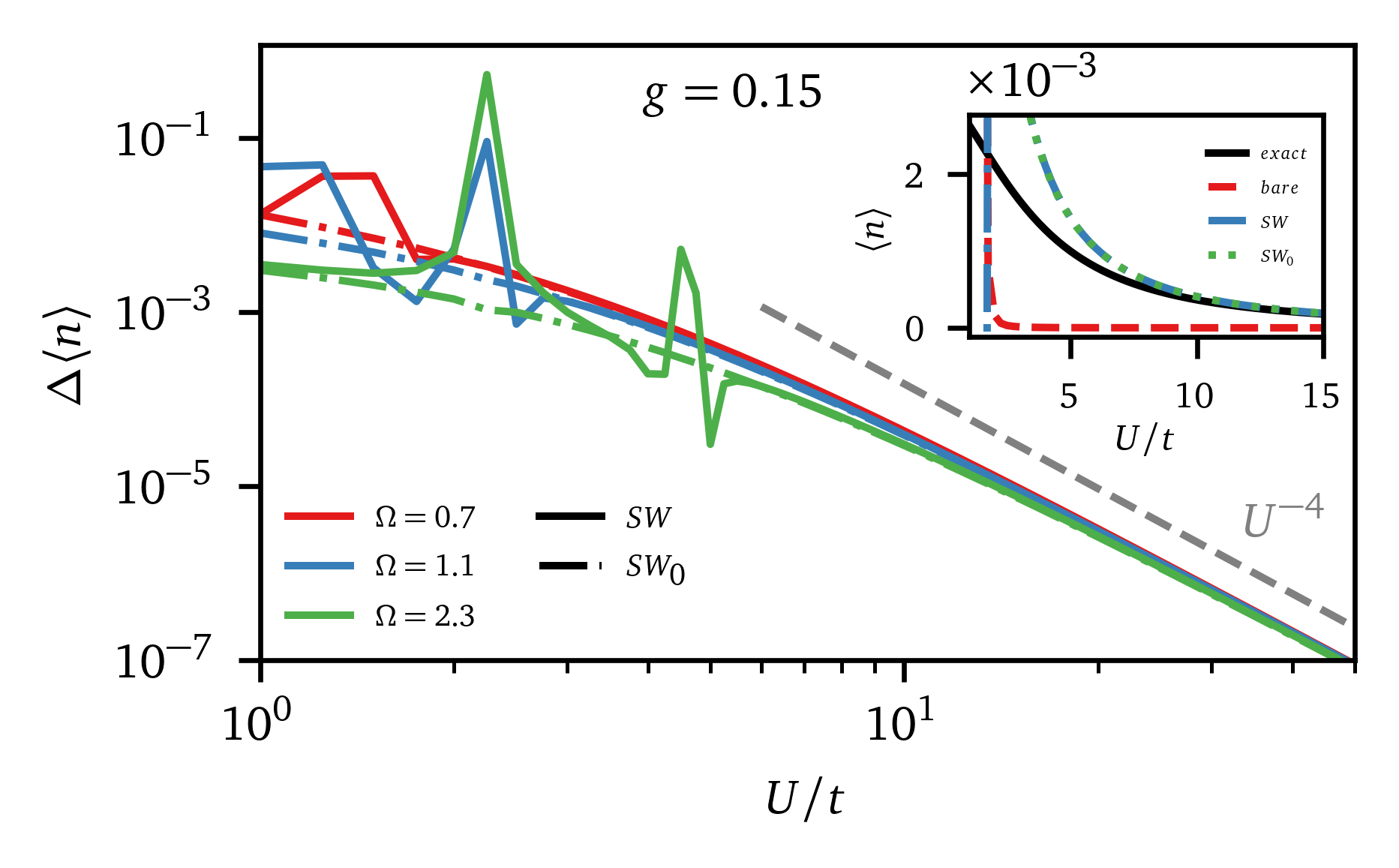}
    \caption{\textbf{Error scaling of SW-corrected occupation $\boldsymbol{\hat{n}}$} using \cref{eq:sw_number_occupation} for $g = 0.15$ and bosonic cutoff $n\_c = 15$ at different cavity frequencies $\Omega$ and spin-orbit interaction $\boldsymbol{\alpha}_{ij} = 0.1 e^{i\pi/4} \mathbb{1}_3$. The main panel shows the absolute error between exact electronic results and the SW corrected observables, while the inset compares the bare data. We observe the expected error scaling of $\order*{t/U}^4$ at large $U$. Next to the full data we also show the SW$_0$ approximation with accuracy comparable to the full solution.}
    \label{fig:sw_error_scaling}
\end{SCfigure}

When taking ground state expectation values of \cref{eq:sw_number_occupation}, terms odd in $S$ will vanish, since $\mathcal{P}S \mathcal{Q} = 0$ and the ground state will lie in the low energy sector. We verify the validity of the formula in \cref{fig:sw_error_scaling}, which shows the error scaling of $\expval{n}$ evaluated in the effective spin model [\cref{eq:cavity_Heff}] compared to the results from the exact electronic model [\cref{eq:cavity_hamiltonian}] for a Hubbard Dimer. While on the operator level, the error scaling of \cref{eq:sw_number_occupation} is $\order*{t/U}^3$, it will be $\order*{t/U}^4$ when taking a simple expectation value due the vanishing odd terms in $S$, as is explicitly shown in the figure.

Next to the data illustrating \cref{eq:sw_number_occupation}, \cref{fig:sw_error_scaling} also shows the SW$_0$ approximation, where we additionally assume that the cavity is perturbatively empty $\ket{\psi} = \ket{\psi\_{m}} \otimes \ket{n = 0} + \order*{t/U}^2$. This amounts to replacing $\sum_{n n'} \dyad{n}{n'} \to \dyad{0}{0}$ in the final expectation value and was used in the main text and in \cite{weberCavityrenormalizedQuantum2023}. It in particular entails that the correction in \cref{eq:sw_number_occupation} involves only a matter expectation value, scaled by photonic constants. This approximation can be motivated perturbatively, by considering the occupation of number states $\ket{n \neq 0}$, viz. $g_{n > 0} = \sum_{n > 0} \norm{(\bra{n} \otimes \mathbb{1} )\ket{\psi}}$. It scales as $\order*{t/U}^2$ for the electronic and as $\order*{t/U}^4$ for the spin model, verifying the validity of the SW$_0$ approximation. This observation further unveils that the leading order contribution of \cref{eq:sw_number_occupation} ($\sim \hat n$) scales as $\expval*{n^{x > 0}}\_{spin} \in \order*{t/U}^4$, making it a subleading contribution in expectation values. We will return to this point when considering correlation functions below. As the Hamiltonian itself, \cref{eq:sw_number_occupation} contains resonances at $U = n\Omega$, where the perturbative expansion breaks down, whose width is controlled by the light-matter interaction strength.

\subsubsection*{Ladder Operators}

The fundamental building block of cavity correlation functions are the ladder operators $a^{(\dagger)}$. These are not directly observables themselves, but their products present observable quantities. Similar to the number operator, one can evaluate the effective ladder operators of the low-energy theory, where one finds after a calculation using \cref{eq:sw_ss_product}
\begin{align}
    \nonumber
    a^\dagger\_{eff} &= a^\dagger + \sum_{nn'} S^{nn'}
    \left( \sqrt{n'} \dyad*{n}{n' - 1} - \sqrt{n + 1} \dyad*{n + 1}{n'} \right)
    + \mathcal{P} \sum_{\expval{i,j}}
    \bigg\{
        (\vb{SS})^+_{ij} \otimes \mathcal{A}^+_{ij} + \vb{S}^-_{ij} \otimes \mathcal{A}^-_{ij}
    \bigg\} \mathcal{P}
    + \mathcal{Q} \dots \mathcal{Q}
    + \order*{t/U}^3 \\
    &= a^\dagger + \delta a^\dagger +
    \sum_{nn'} S^{nn'} \left( \sqrt{n'} \dyad*{n}{n' - 1} - \sqrt{n + 1} \dyad*{n + 1}{n'} \right)
    + \order*{t/U}^3,
    \label{eq:sw_ladder}
\end{align}
with the fairly complicated photonic correction operators
\begin{align}
    \mathcal{A}^\pm_{ij}
    = \frac{U}{4} \sum_{nn'} \sum_k
    \Biggl\{
    &\Phi^{\pm, nk k(n' + 1)}_{ij}
    \frac{\sqrt{n' + 1}}{[U + \Omega (k - n)] [U + \Omega (k - (n' + 1))]}
    \\ \nonumber
     + &\Phi^{\pm, (n - 1)k kn}_{ij}
    \frac{\sqrt{n}}{[U + \Omega (k - (n - 1))] [U + \Omega (k - n')]}
    \\ \nonumber
    -2 &\Phi^{\pm, n(k + 1) kn}_{ij}
    \frac{\sqrt{k + 1}}{[U + \Omega ((k + 1) - n)] [U + \Omega (k - n')]}
    \Biggr\} \dyad*{n}{n'},
    \label{eq:sw_ladder_photon_correction}
\end{align}
and the shorthand $\delta a^\dagger = \sum_{\expval{i,j}} \left\{ (\vb{SS})^+_{ij} \otimes \mathcal{A}^+_{ij} + \vb{S}^-_{ij} \otimes \mathcal{A}^-_{ij} \right\}$. Here the first two terms correspond to processes where a photon is created in the low energy sector, which can be made explicit by combining them to $ \phi^{\pm, nkkn'} ... (\sqrt{n'} \dyad*{n}{n'-1} + \sqrt{n + 1} \dyad*{n + 1}{n'})$ by renaming indices. The last term describes the processes involving the creation of a photon in the high energy sector instead.

\subsection{Dynamical Correlation functions}

Next to static observables, dynamical correlation functions play an important role for probing properties of matter. In this section we will illustrate that dynamical photonic correlation functions $C(t) = \expval*{\mathcal{O}_1(t) \mathcal{O}_2}$ are related to dynamical matter correlations, and contain information about the selection rule resolved excitation spectrum.

Before proceeding, we comment on the time dependence of SW transformed observables. Since the SW transform is a unitary transformation, it will act independently on the different contributions of the time evolution
\begin{align}
    \mel{\psi}{e^{iHt} A e^{-iHt}}{\psi} &=
    \bra{\phi}
    \left[ e^S e^{-iHt} e^{-S} \right]
    \left[ e^S A e^{-S} \right]
    \left[ e^S e^{iHt} e^{-S} \right]
    \ket{\phi}
    =
    \bra{\phi}
    e^{-i\tilde{H}t} \left[ e^S A e^{-S} \right] e^{i\tilde{H}t}
    \ket{\phi},
\end{align}
with $\ket{\psi}$ the exact ground state. $\tilde{H} = e^S H e^{-S}$ and $\ket{\phi} = e^S \ket{\psi}$ denote the SW transformed state and Hamiltonian respectively. The effective low-energy Hamiltonian $\tilde{H}$ can be replaced with the effective Hamiltonian $H\_{eff} = \mathcal{P} \tilde{H} \mathcal{P}$, since $\mathcal{P}\tilde{H}\mathcal{Q} = 0$ and $\ket{\phi} \in \mathcal{P}$ by definition of the ground state. We proceed by introducing the (regularized) Laplace transform of the correlation function also defined in the main text
\begin{align}
    C(\omega) =
    \int_0^\infty \dd{t} e^{i(\omega + i\eta)t} \mel{\phi}{e^{iHt} \mathcal{O}_1 e^{-iHt}\mathcal{O}_2  }{\phi}
    = i \mel{\phi}{\mathcal{O}_1 \frac{1}{\omega + i\eta - (H - E_\phi)} \mathcal{O}_2}{\phi},
    \label{eq:corr_laplace}
\end{align}
where we assumed $\ket{\phi}$ to be an eigenstate of $H$ with energy $E_\phi$. \Cref{eq:corr_laplace} takes the form of a resolvent operator measured in $\mathcal{O}_{1,2} \ket{\phi}$, hence probing energy and lifetime of the excitation created by the operator(s).

\subsubsection{Number-Number Correlator}

We consider the intensity-intensity correlation function $N_2(t) = \expval*{n n(t)_H}$,
which we can directly evaluate using \cref{eq:sw_number_occupation}
\begin{align}
    \nonumber
    N_{2, \textnormal{eff}}(t) &=
    \expval*{n~n(t)_{H\_{eff}}} +
    \expval*{\delta n ~ n(t)_{H\_{eff}}} +
    \expval*{n ~ \delta n(t)_{H\_{eff}}} +
    \expval*{\delta n ~ \delta n(t)_{H\_{eff}}} + \\
    &+ \sum_{nn'} \sum_{mm'} (n' - n) (m' - m)
    \expval{
        \bigl[ S^{nn'} \otimes \dyad*{n}{n'} \bigr]
        \bigl[ S^{mm'} \otimes \dyad*{m}{m'} \bigr](t)_{H\_{eff}}
    }
    + \order*{t/U}^3.
    \label{eq:corr_N2_time}
\end{align}
This expression contains two types of contributions. The first row, describes low energy magnetic and photonic excitations, while the second row, involving a product of SW operators, describes high energy doublon-polariton quasi-particles. A more concise representation of \cref{eq:corr_N2_time} is given in frequency space [\cref{eq:corr_laplace}]
\begin{align}
    \nonumber
    N\_{2,eff}(\omega) &= N\_{2,doub.}(\omega) + N\_{2,low}(\omega) \\
    &=
    N\_{2,doub.} +
    i \mel{\phi}{
    \left(n + \delta n \right)
    \frac{1}{\omega + i\eta + H - E_\phi}
    \left(n + \delta n \right)}{\phi}.
    \label{eq:corr_N2_w}
\end{align}
The high energy term $N\_{2,doub.}$ gives the leading order contribution to the time evolution [$\order*{t/U}^2$] and can be explicitly evaluated using the projection structure of $S$ and $\tilde{H}$, viz. $\mathcal{P} \tilde{H} \mathcal{Q} = 0$ and $\mathcal{P} S \mathcal{P} = 0$. Since $\mathcal{P} \ket{\phi} = \ket{\phi}$ by definition of the ground state, the time evolution operator in between the two SW operators only acts on the high energy-sector [$\dots \mathcal{Q} e^{i\tilde{H}t} \mathcal{Q} \dots$] and can be evaluated explicitly to find
\begin{align}
    \nonumber
    N\_{2,doub.}(\omega) &=
    \sum_{nn'} \sum_k \frac{i}{\omega + i\eta - (U + \Omega(k + 1/2) - E_\phi)} \\
    & \hspace{2cm}
    \times \frac{U}{2} \sum_{\expval{i,j}} \frac{-(k - n) (k - n')}{[U + \Omega (k - n)] [U + \Omega (k - n')]}
    \expval{
    \left(  (\vb{SS})^+_{ij} \Phi^{+, nkkn'}_{ij} + \vb{S}^-_{ij} \Phi^{-, nkkn'}_{ij} \right)
    \otimes \dyad*{n}{n'}},
    \label{eq:corr_number_high_energy}
\end{align}
which describes doublon-polariton excitation with energy $U + \Omega (k + 1/2)$ [see term in $\omega$-dependent fraction], that are generated by $\mathcal{Q} S \mathcal{P} \cdot a^\dagger a$ acting on the ground state. While this contribution perturbatively gives the leading order, the excitation are high energy $\sim U$ and hence not of interest when studying the low energy dynamics of the model, where $N\_{2,doub.} \approx const$. This is instead governed by the second term in \cref{eq:corr_N2_w}.

At first glance it might seem inconsistent to retain the contribution $\delta n \cdot \delta n$, but in fact it is not! All low-energy terms in \cref{eq:corr_N2_w} contribute at the same order of perturbation theory, i.e. $\order*{t/U}^4$. To understand this, note that we earlier observed that $\expval*{n^x}\_{spin} \sim \order{t/U}^4$, which results from the SW$_0$ approximation $\ket{\psi} = \ket{\psi\_e} \otimes \ket{n = 0} + \order*{t/U}^2$. Hence, all terms $\sim n$ get suppressed by that factor. Further, this expansion is consistent, since there are no other terms that can contribute at $\order*{t/U}^4$.

We proceed by analyzing \cref{eq:corr_N2_w} numerically for the Hubbard plaque model (without LS interactions), considered in the main text, viz. a model with 4 lattice sites arranged on a square 4 with an electromagnetic field $\vb{A} = A\vb{e_x}$, resulting in photon dressing of hoppings in $x$ direction only. This inhomogeneity of the light-matter coupling allows us to probe the selection rule allowed singlet excitation of the model with $\Delta S, \Delta m = 0$ (further explained below).

\begin{figure}[b]
    \centering
    \includegraphics[width = \linewidth]{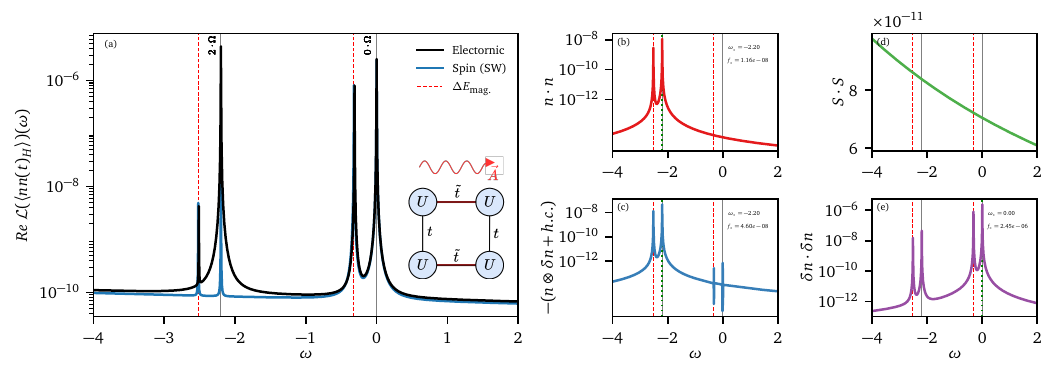}
    \caption{\textbf{Photonic $\boldsymbol{N_2}(\omega)$ correlation function for the Hubbard Plaque} at $U = 25, \Omega = 1.1, \boldsymbol{\alpha}_{ij} = 0, g = 0.1$ and $\eta = 10^{-3}$.
    Vertical gray lines indicate cavity resonances at $n \cdot \Omega$, while the red dashes indicate the selection rule allowed low energy excitation of the matter, which appear as a bare resonance at $\Delta E$ and a polariton-spin excitation at $\Delta E + 1 \cdot \Omega$. (a) Exact electronic frequency space representation of $N_2(\omega)$ (black) and approximate low-energy expression $N\_{2,eff}(\omega)$ [\cref{eq:corr_N2_w}] for the spin model (blue). Panels (b) - (e) show the contribution resolved $N\_{2,eff}(\omega)$, illustrating the importance of the different terms. Expectation values evaluated in full spin ground state, since SW$_0$ leads to no significant changes.}
    \label{fig:correlations_n2_plaque}
\end{figure}

\Cref{fig:correlations_n2_plaque} illustrates this by example and shows the numerical results for $N_2(\omega)$ obtained from full electronic (black) and low-energy spin theory (blue), which agree well in the low energy sector but shows stronger deviations at higher energies. Despite this, the approximate treatment reproduces the position of the excitation peaks with excellent precision. Vertical gray lines indicate cavity resonances at $n \cdot \Omega$, while the red dashes indicate the selection rule allowed low energy excitation of the matter at $\Delta E$, and a polariton-spin excitation at $\Delta E + 1 \cdot \Omega$. Considering the resolved results, shown on the right hand side of \cref{fig:correlations_n2_plaque}, we observe that the dimer-dimer term $\delta n \cdot \delta n$ is the dominant contribution by two orders of magnitude. This remains true for the full range of considered parameters, and hints at the connection between photonic observable and spin dimer-dimer correlation functions which we can make explicit by additionally considering a perturbative expansion in the light-matter interaction.

\begin{SCfigure}[][h]
    \centering
    \includegraphics{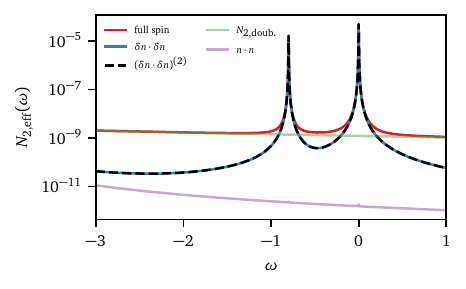}
    \caption{\textbf{Full and perturbative photonic correlation function $\boldsymbol{N_2(\omega)}$} at $g = 0.15, \Omega = 2.34$ and $U = 10$. $N\_{2,eff.}$ [\cref{eq:corr_N2_w}] (red) is compared with the $\delta n \cdot \delta n$ contribution (blue) and it's perturbative approximation \cref{eq:corr_N2_perturbative_w} (black-dashed), which are in excellent agreement, highlighting that $N_2$ measures the specified matter correlator, only missing the (trivial) structure-less background provided by $N\_{2,doub,}$ (green desat.) and bare contribution $n \cdot n$ (purple desat.). }
    \label{fig:corr_N2_perturbative_w}
\end{SCfigure}

Inspecting \cref{eq:sw_number_photon_correction}, we note that the perturbative expansion in $g$ amounts to expanding the symmetrized product of Peirls phase factors which leads to the perturbative expansion for the number operator correction factor
\begin{align}
    \label{eq:N_ij_perturbative}
    \mathcal{N}_{ij}^+ &= - \frac{g_{ij}^2}{2N} \sum_{n n'}
    \left\{
        (x^2)^{nn'} \frac{\Omega (n - n')}{U^2 - \Omega^2(n - n')^2} -
        \frac{2}{U} \sum_k x^{nk} x^{kn'} \frac{\frac{1}{2}(n + n') - k}{[U + \Omega (k - n)] [U + \Omega (k - n')]}
    \right\} \dyad*{n}{n'} + \order{g}^4, \\ \nonumber
    &= - \frac{g_{ij}^2}{N} \sum_{nn'} \delta \mathcal{N}^{(2)}_{nn'} \dyad*{n}{n'} + \order{g}^4,
\end{align}
with $g_{ij}$ defined below \cref{eq:cavity_hamiltonian}. Note in particular that there is no constant in $g$ contribution to the number operator correction. Within the SW$_0$ approximation the wave function takes the form $\ket{\psi} = \ket{\psi\_e} \otimes \ket{n = 0} + \order*{t/U}^2$, so that matter and photonic contributions factorize. In particular when considering the $(\delta n \cdot \delta n) (\omega)$, we are left with a matter correlation function multiplied with a constant depending on model parameters. To see this consider
\begin{align}
    (\delta n \cdot \delta n) (\omega) = \mel{\psi\_e}{\delta n (\omega + i \eta + H - E_\phi)^{-1} \delta n}{\psi\_e} =
    \mel{\psi\_e}{\delta n^{(2)} (\omega + i \eta + H\_e - E\^e_\phi)^{-1} \delta n^{(2)}}{\psi\_e} + \order{g}^4,
\end{align}
with the LM decoupled ($g = 0$) Hamiltonian $H\_e$ and ground state energy $E\^e_\phi$. Inserting \cref{eq:sw_number_occupation} and using \cref{eq:N_ij_perturbative} we observe that $N_2(\omega)$ directly probes the dimer-dimer correlation function
\begin{align}
    (\delta n \cdot \delta n) (\omega) =
    \mel{\psi\_e}{
    \sum_{\expval{i,j}} \frac{g_{ij}^2}{N} J_{ij} \left( \vb{S}_i \cdot \vb{S}_j - 1/4 \right)
    \frac{i}{\omega + i \eta + H\_e - E\^e_\phi}
    \sum_{\expval{k,l}} \frac{g_{kl}^2}{N} J_{kl} \left( \vb{S}_k \cdot \vb{S}_l - 1/4 \right)
    }{\psi\_e} \cdot c(U, \Omega)
    \label{eq:corr_N2_perturbative_w}
\end{align}
with the photonic constant $c(U, \Omega) =  \sum_m \delta \mathcal{N}_{0, m}^{(2)} \delta \mathcal{N}_{m, 0}^{(2)} \in \mathbb{R}$, depending only on model constants, but not on the photonic ground state itself. Hence, one can write for the low energy dynamics
\begin{align}
    \expval{n \; n(t)}\_{eff} =
    c(U, \Omega)
    \expval{
        \left[\sum_{\expval{1,2}} \frac{g_{12}^2}{N} J_{12} \left( \vb{S}_1 \cdot \vb{S}_2 - 1/4 \right) \right] \;
        \left[\sum_{\expval{3,4}} \frac{g_{34}^2}{N} J_{34} \left( \vb{S}_3 \cdot \vb{S}_4 - 1/4 \right) \right](t)
        } + N_{2, \rm{doub.}}(t) + \order{g}^4.
        \label{eq:corr_N2_perturbative_t}
\end{align}
The low-energy dynamics, governed by the dimer-dimer correlation function in \cref{eq:corr_N2_perturbative_t}, is superimposed by fast, high-energy oscillations $N_{2, \rm{doub.}}(t)$ due to doublon-polariton excitations, which occur at a higher energy scale and are less prominent in shaping the overall behavior. In particular, their contribution in frequency space is approximately constant at low energies $N_{2, \rm{doub.}}(\omega) \approx const$ [\cref{fig:corr_N2_perturbative_w} green]. Consequently, we focus on the low-energy dynamics moving forward.

We note that for uniform LM interactions, viz. $g_{ij} \equiv g$, the dimer operators in \cref{eq:corr_N2_perturbative_t} commute with the Hamiltonian, so that we do not measure any non-trivial low energy dynamics, explaining why the non-uniform light matter coupling was necessary in the Hubbard plaque above. Without LS interactions, a uniform light-matter coupling does not probe matter excitations. But for non-uniform LM coupling topologies, e.g. the one found for the Hubbard plaque, the $N_2(t)$ correlation function directly probes a matter correlator, viz. a certain dimer-dimer correlation function. By engineering the LM coupling topology we can control which correlation function is measured.

We investigate \cref{eq:corr_N2_perturbative_w} for the Hubbard plaque in more detail in \cref{fig:corr_N2_perturbative_w}. In the figure $N\_{2,eff.}$ [\cref{eq:corr_N2_w}] (red) is compared with the $\delta n \cdot \delta n$ contribution (blue) and it's perturbative approximation \cref{eq:corr_N2_perturbative_w} (black-dashed), which are in excellent agreement, highlighting that $N_2$ measures the specified matter correlator \cref{eq:corr_N2_perturbative_t}, only missing the (trivial) structure-less background provided by $N\_{2,doub,}$ (green) and bare contribution $n \cdot n$ (purple). These contributions are of course only structure-less in the low energy sector, but since this is the energy range of interest, using the perturbative expansion of $\delta n \cdot \delta n$ constitutes a valid approximation to the full signal. We hence provided an explicit expression for the matter correlation function probed by measuring $N_2(\omega)$.

\newpage
\subsubsection{Coherence functions}

Next to the intensity-intensity correlation function, we analyze the coherence functions of the photonic field $C(t) = \expval*{a^\dagger a(t)_H}$ in this section and show, that it also contains similar information about the matter spectrum, in particular also probes the selection rule allowed magnetic excitation. The analytical expressions found are considerably more complicated making the accurate interpretation more challenging. The SW transformation of $C(t)$ can be directly obtained using \cref{eq:sw_ladder} to find
\begin{align}
    \label{eq:corr_coherence_time}
    C\_{eff}(t) &= \expval{a^\dagger~ a(t)_{H\_{eff}}}
    + \expval{a^\dagger~ \delta a(t)_{H\_{eff}}}
    + \expval{\delta a^\dagger~ a(t)_{H\_{eff}}}
    + \expval{\delta a^\dagger~ \delta a(t)_{H\_{eff}}} \\
    &+ \sum_{nn'}\sum_{mm'}
    \Big[
        S^{nn'} \left( \sqrt{n'} \dyad*{n}{n' - 1} - \sqrt{n + 1} \dyad*{n + 1}{n'} \right)
    \Big]
    \Big[
        S^{mm'} \left( \sqrt{m' + 1} \dyad*{m}{m + 1} - \sqrt{m} \dyad*{m-1}{m'} \right)
    \Big](t)_{H\_{eff}}
    \nonumber
\end{align}
or again more concisely in frequency space [see \cref{eq:corr_laplace}]
\begin{align}
    C(\omega) = C\_{doub.}(\omega) +
    i \mel{\phi}{
    \left(a^\dagger + \delta a^\dagger \right)
    \frac{1}{\omega + i\eta + H - E_\phi}
    \big(a + \delta a \big)}{\phi}.
    \label{eq:corr_coherence_w}
\end{align}
Notice in particular the similar structure to \cref{eq:corr_N2_w}, leading to the separation of low ($\sim t / U$) and high ($\sim U$) energy contributions. For the latter, one can again explicitly derive an explicit expression, describing the doublon-polariton excitations of the system. One finds the lengthy expression
\begin{align}
    C\_{doub.}(\omega) &=
    \sum_{nn'} \sum_k \frac{i}{\omega + i\eta - (U + \Omega(k + 1/2) - E_\phi)} \sum_{\expval{i,j}}
        \frac{U}{2[U + \Omega (k + 1 - n)] [U + \Omega (k + 1 - n')]}
    \\
    & \hspace{1cm} \times \Bigg\langle \Bigg\{
    \sqrt{k + 1} \sqrt{n'}\phantom{+1}
    \left(
        (\vb{SS})^+_{ij} \Phi^{+, n(k+1)k(n'-1)}_{ij} + \vb{S}^-_{ij} \Phi^{-,  n(k+1)k(n'-1)}_{ij}
    \right) \\
    & \hspace{1.5cm} +
    \sqrt{k + 1} \sqrt{n}\phantom{+1'}
    \left(
        (\vb{SS})^+_{ij} \Phi^{+,(n-1)k(k+1)n'}_{ij} + \vb{S}^-_{ij} \Phi^{-,(n-1)k(k+1)n'}_{ij}
    \right) \\
    & \hspace{1.5cm}-
    \sqrt{k + 1}\sqrt{k + 1}
    \left(
        (\vb{SS})^+_{ij} \Phi^{+,n(k+1)(k+1)n'}_{ij} + \vb{S}^-_{ij} \Phi^{-,n(k+1)(k+1)n'}_{ij}
    \right) \\
    & \hspace{1.5cm}-
    \sqrt{n}\sqrt{n'} \phantom{+1+1'}
    \left(
        (\vb{SS})^+_{ij} \Phi^{+, (n-1)kk(n'-1)}_{ij} + \vb{S}^-_{ij} \Phi^{-,(n-1)kk(n'-1)}_{ij}
    \right)
    \Bigg\} \otimes \dyad*{n}{n'} \Bigg\rangle.
    \label{eq:corr_coherence_high_energy}
\end{align}
Despite being the leading order in perturbation theory, this term is of subleading importance for understanding the low energy dynamics of the model, where it only provides an approximately constant background. The dominant low-energy contribution is again given by the second term in \cref{eq:corr_coherence_w}. The leading order contribution for the low-energy part scales as $\order*{t/U}^4$, which follows from the previously established $\expval*{n^{x > 0}} = \order*{t/U}^4$ and SW$_0$ approximation. Further, since $\delta a^{(\dagger)}$ also gives a finite contribution, when the cavity is empty, all low-energy terms give contributions of the same perturbative scaling, viz. $\order*{t/U}^4$.

\begin{figure}[t]
    \centering
    \includegraphics[width = \linewidth]{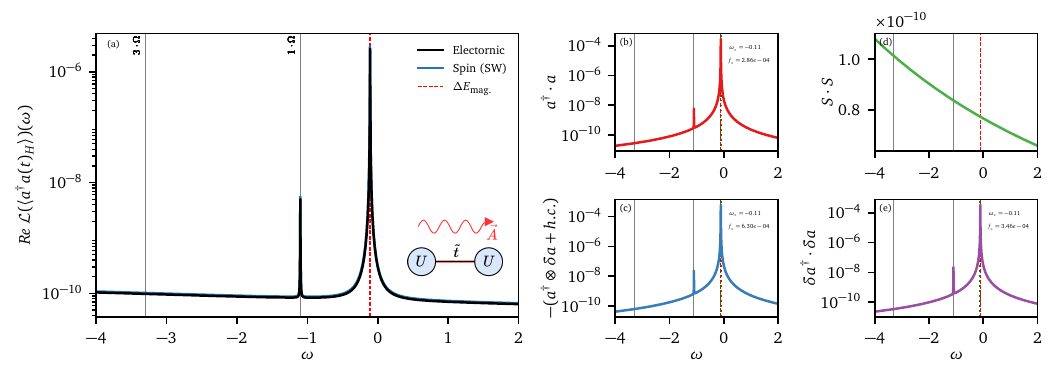}
    \caption{\textbf{Photonic coherence function $\boldsymbol{C}(\omega)$ for the Hubbard dimer} at $U = 25, \Omega = 1.1, \boldsymbol{\alpha}_{ij} = 0.75 \mathbb{1}, g = 0.1$ and $\eta = 10^{-3}$. Vertical gray lines indicate cavity resonances at $n \cdot \Omega$, while the red dashes line shows the selection rule allowed low energy excitation of the matter. (a) Exact electronic frequency space representation of $N_2(\omega)$ (black) and approximate low-energy expression derived in \cref{eq:corr_N2_w} for the spin model (blue), which is mostly hidden by the exact data. Panels (b) - (e) show the contribution resolved \cref{eq:corr_N2_w}, illustrating the importance of the different terms. Spin expectation values evaluated within full spin ground state, since SW$_0$ leads to no significant changes.}
    \label{fig:correlations_coherence_dimer}
\end{figure}

As for the intensity-intensity correlation, studying the photonic coherence functions allows us to extract information about the selection rule filtered matter excitations of the system. The Hubbard dimer only hosts triplet low energy excitations with $\Delta S, \Delta m = 0$, that can't be probed by photons in the current context [see \cref{eq:corr_N2_perturbative_t}]. Including strong LS interactions this can be alleviated, revealing a low energy magnetic excitation in the photonic signal as we show in \cref{fig:correlations_coherence_dimer}, which illustrates $C(\omega)$ obtained from exact electronic (black) and approximate spin theory (blue). The height of the magnetic excitation peak (red dashed) scales with the LS interaction strength and vanishes for $|\boldsymbol{\alpha}| \to 0$.

Panels (b) - (e) show the contribution resolved $C\_{eff}(\omega)$, with panel (d) depicting the high-energy contribution which is structure-less for low energies. The other panels depict the contributions to the low-energy dynamics, which are all on the same order of magnitude and contain very similar spectral information. Since the different terms contribute at the same order of magnitude, the identification of the photonic correlation function with the "dimer-dimer" correlation function $\delta a^\dagger \cdot \delta a$ becomes questionable. Nonetheless, the dynamical photon correlation function can be used to probe the selection rule filtered excitation spectrum.

\bibliographystyle{apsrev4-1}
\bibliography{bibliography.bib}